\documentclass[modern]{aastex62}

\usepackage[sort&compress]{natbib}
\usepackage{graphicx}
\usepackage{xspace}
\usepackage{xcolor}
\usepackage{bm}
\usepackage{mathtools}
\shorttitle{a continuous correlation function estimator}
\shortauthors{storey-fisher and hogg}

\newcommand{\cf}{2pcf\xspace}
\newcommand{\Est}{The Continuous-Function Estimator\xspace}
\newcommand{\est}{the Continuous-Function Estimator\xspace}
\newcommand{\documentname}{\textsl{Article}\xspace}
\newcommand{\LS}{LS\xspace}

\newcommand{\eqt}[1]{equation~(\ref{#1})}

\newcommand{\inv}{^{-1}}
\newcommand{\invp}{^{'-1}}
\newcommand{\T}{^{\mathsf{T}}}
\newcommand{\Tp}{^{'\mathsf{T}}}
\newcommand{\hmpc}{$h^{-1}\,$Mpc}
\newcommand{\dd}{\mathrm{d}}
\newcommand{\bld}[1]{\bm{#1}}
\newcommand{\vv}[1]{\bld{v}_\mathrm{#1}}
\newcommand{\TT}[1]{\bld{T}_\mathrm{#1}}
\newcommand{\ff}{\bld{f}}
\newcommand{\NN}[1]{N_\mathrm{#1}}
\newcommand{\GG}[1]{\mathsf{G}_{#1}}

\newcommand{\new}[1]{#1}

\newcommand{\ccpp}{\affiliation{%
    Center for Cosmology and Particle Physics,
    Department of Physics,
    New York University}}
\newcommand{\flatiron}{\affiliation{%
    Flatiron Institute, Simons Foundation}}
\newcommand{\cds}{\affiliation{%
    Center for Data Science,
    New York University}}
\newcommand{\mpia}{\affiliation{%
    Max-Planck-Institut f\"{u}r Astronomie, Heidelberg}}

\begin{document}\sloppy\sloppypar\raggedbottom\frenchspacing

\title{\textbf{Two-point statistics without bins: A continuous-function generalization of the correlation function estimator for large-scale structure}}

\author[0000-0001-8764-7103]{Kate Storey-Fisher}
\ccpp
\correspondingauthor{Kate Storey-Fisher \texttt{<\href{mailto:k.sf@nyu.edu}{k.sf@nyu.edu}>}}

\author[0000-0003-2866-9403]{David W. Hogg}
\ccpp
\cds
\mpia
\flatiron

\begin{abstract}\noindent
The two-point correlation function (2pcf) is the key statistic in structure formation; it measures the clustering of galaxies or other density field tracers. Estimators of the 2pcf, including the standard Landy--Szalay (LS) estimator, evaluate the 2pcf in hard-edged separation bins, which is scientifically inappropriate and results in a poor trade-off between bias and variance. We present a new 2pcf estimator, the Continuous-Function Estimator, which generalizes LS to a continuous representation and obviates binning in separation or any other pair property. Our estimator, inspired by the mathematics of least-squares fitting, replaces binned pair counts with projections onto basis functions; it outputs the best linear combination of basis functions to describe the 2pcf. The choice of basis can take into account the expected form of the 2pcf, as well as its dependence on pair properties other than separation. We show that the Continuous-Function Estimator with a cubic-spline basis better represents the shape of the 2pcf compared to LS. We also estimate directly the baryon acoustic scale, using a small number of physically-motivated basis functions. Critically, this leads to a reduction in the number of mock catalogs required for covariance estimation, which is currently the limiting step in many 2pcf analyses. We discuss further applications of the Continuous-Function Estimator, including determination of the dependence of clustering on galaxy properties and searches for potential inhomogeneities or anisotropies in large-scale structure.
\end{abstract}


\section{Introduction}

The large-scale structure (LSS) of the Universe is critical to our understanding of fundamental cosmology. 
It encodes information about the physics of the early Universe and the subsequent expansion history \citep{SunyaevZeldovich1970, HuSugiyama1996, Riess1998}.
In particular, LSS measures the Baryon Acoustic Oscillation (BAO) scale \citep{Cole2005, Eisenstein2005}, which results from density fluctuations in the baryon--photon fluid.
The distance traveled by these density waves before recombination imprints a feature on the statistical description of the LSS, which can be used to determine the characteristic BAO length scale \citep{PeeblesYu1970, EisensteinHu1998}.
The LSS also contains the signature of redshift-space distortions caused by the peculiar velocities of galaxies, which are used to measure the growth rate of structure \citep{Kaiser1987}.
Additionally, the LSS can be used to constrain galaxy formation in conjunction with models of galaxy bias (e.g., \citealt{Hamilton1988},  \citealt{Li2006}, \citealt{Zehavi2011}, \citealt{Durkalec2018}).
With current observations, the LSS is well-described by a cold dark matter model with a cosmological constant, the standard $\Lambda$CDM model (e.g., \citealt{Alam2016}).
Upcoming galaxy surveys including DESI \citep{Aghamousa2016}, Euclid \citep{Laureijs2011}, and LSST \citep{Ivezic2018} will observe larger volumes with improved measurements, allowing us to test $\Lambda$CDM to even higher precision.

The most important statistic for characterizing the LSS is the two-point correlation function (\cf).
\new{It measures the excess probability of finding two galaxies are separated by a given distance, compared to a spatially random Poisson distribution;} effectively, it characterizes the strength of clustering at a given spatial scale.
The \cf is the primary tool for extracting cosmological information from galaxy redshift surveys.
Such correlation function analyses include \cite{Hawkins2003} for the 2dF Galaxy Redshift Survey (2dFGRS, \citealt{Colless2001}), \cite{Alam2016} for the Baryon Oscillation Spectroscopic Survey (BOSS, \citealt{Dawson2013}) DR12 analysis, and \cite{Elvin-Poole2017} for the Dark Energy Survey (DES, \citealt{DES2005}).

Traditionally, the \cf is estimated in bins of radial separation.
\new{Recent work has focused on the inappropriateness and limitations of binning in astrophysical contexts.}
Broadly, binning adds arbitrary boundaries between continuous data; results should not depend on bin choice, yet they sometimes do.
\new{It results in the well-known trade-off between bias and variance: fewer bins may bias the result (i.e. reduce the accuracy), while more bins will increase the variance of measurement (i.e. reduce the precision)}.
Finite-width bins also result in a loss of information about the property in which one is binning.
\new{These issues have been demonstrated in various fields of astrophysics.}
\new{\cite{Kipping2010} showed that temporal binning of transit lightcurves results in inaccurately recovered system parameters.}
\cite{Lanzuisi2017} noted that the choice of binning axis impacts the detected correlation between the luminosity of active galactic nuclei and their host galaxies; \cite{Grimmett2020} devised a method to investigate this correlation in a continuous manner using a hierarchical Bayesian model, eliminating the need for binning.
\new{\cite{Brout2020} found that binning supernovae data causes a larger systematic error in inferred cosmological parameters, while performing an unbinned analysis allows the data to self-calibrate out these systematic uncertainties.}
From this literature it is clear that, when analyzing smooth quantities, binning is sinning.

\new{The issue of binning is particularly relevant in LSS analyses.}
\cite{Bailoni2016} explored the dependence of clustering analyses on the number of redshift bins, finding a non-negligible difference in cosmological parameter uncertainties.
\new{\cite{Schneider2009} showed that in cosmic shear analyses, increasing the number of bins increases the probability that the correlation function is not positive semi-definite, and therefore not statistically valid.}
The implications for BAO analyses were explored by \cite{Percival2014}, who found that the effects of bin width are small but non-negligible; they showed that there is an optimal bin width given the analysis method that trades off statistical uncertainty against bias in the derived BAO peak location.
Generally, the loss of information inherent in binning may become a critical bottleneck as we work towards extreme precision in LSS analyses.

Another critical issue for \cf analyses is that the error on the inverse covariance matrix estimate depends on the number of bins.
A larger number of bins results in a larger error that propagates to the estimated parameters \citep{Dodelson2013}.
This can be balanced by using a large number of mock galaxy catalogs, but these can be exceedingly expensive to generate.
Covariance matrix estimation is currently the limiting step in many LSS analyses.
For galaxy clustering, on the order of 1000 mock catalogs tailored to the survey needed to achieve the desired precision on the parameters \citep{Percival2014}.
\new{Weak lensing analyses especially suffer from this issue, due to the very high-dimensional data vectors (for a review see \citealt{Mandelbaum2018a}).}
As survey size increases and we work towards higher precision, the requirements on the covariance matrix will get even more stringent; the connection of this limiting step with bin choice merits scrutiny of binning in \cf analyses.

Estimators of the \cf have been studied extensively (e.g., \citealt{PeeblesHauser1974}; \citealt{DavisPeebles1983}; \citealt{Hamilton1993}).
One of the difficulties in performing a two-point estimate is that nontrivial survey boundaries would bias a direct summation of pair counts.
To account for the boundaries as well as \new{corrupted regions (e.g. by bright foreground stars)}, typically a large set of random points are \new{Poisson-distributed} within the acceptable survey window.
The pairwise correlations of these unclustered points are used to normalize out the survey window.
The current standard estimator, proposed by \cite{LandySzalay1993} (hereafter \LS), takes this approach.
It involves a summation of the data--data pairs $DD$ in each separation bin, \new{and uses the random--random pairs $RR$ to perform this edge-correcton;} the data--random pairs $DR$ are incorporated to improve the bias and variance properties of the estimator \new{(see e.g. \citealt{Hamilton1993})}.
The \LS estimator of the correlation function $\hat{\xi}_k$ for the $k^\mathrm{th}$ bin in separation $r$ is defined as
\begin{equation} \label{eq:lsintro}
\hat{\xi}_k = \frac{DD_k - 2DR_k + RR_k}{RR_k}
\end{equation}
\new{where we have assumed the binned pair counts are normalized by the total number of pairs for each set of catalogs}.
Compared with other estimators based on simple combinations of $DD$, $DR$ and $RR$, \LS has been shown to have the lowest bias and variance \citep{Kerscher2000}.
Estimators of the \cf must also take into account the imperfect nature of the survey, including systematic effects, the target completeness, and fiber collisions.
To account for these, each galaxy pair is typically assigned a weight, and pair counts are replaced by the sum of pair weights.

Variations on traditional \cf estimation have been proposed in recent years.
\cite{Demina2016} replaced the $DR$ and $RR$ terms with an integral over the probability map, reducing computation time and increasing precision.
An estimator proposed by \cite{VargasMagana2013} iterates over sets of mock catalogs to find an optimal linear combination of data and random pair counts, reducing the bias and variance.
An alternative estimator, the marked correlation function (e.g., \citealt{WhitePadmanabhan2009}), avoids the use of a random catalog altogether: it considers the ratio between the \cf and a weighted correlation function in which weights are assigned based on galaxy properties, such as the local density.
These estimators have all taken probabilistic approaches; others have taken a likelihood approach.
\cite{BaxterRozo2013} introduced a maximum likelihood estimator for the \cf, which achieves lower variance compared to the \LS estimator, enabling finer binning and requiring a smaller random catalog for the same precision.

These estimators present improvements to \LS, but they are still limited to estimates in separation bins.
Some require additional computational costs or layers of complexity, so the standard formulation of \LS continues to be the default estimator used in most analyses.

In this \documentname, we present a new estimator for the correlation function, \est, which generalizes the \LS estimator to produce a continuous estimation of the \cf. 
\Est projects the galaxy pairs onto a set of continuous basis functions and directly computes the best-fit linear combination of these functions.
The basis representation can depend on the pair separation as well as other desired properties, and can utilize the known form of the \cf.
For tophat basis functions, the estimator exactly reduces to the \LS estimator. 
\Est removes the need for binning and produces a more representative estimate of the \cf with fewer basis functions, \new{increasing the accuracy and} reducing requirements on mock catalogs for covariance matrix computation.
It is particularly well-suited to the analysis of LSS features such as the BAO peak; we find that we can accurately locate the peak with fewer components compared to standard analyses.

This \documentname is organized as follows. 
In Section~\ref{sec:motiv}, we motivate our estimator and explain its formulation, \new{and derive the connection to least-squares fitting}.
We demonstrate its application on a simulated dataset, including a toy BAO analysis, in Section~\ref{sec:experiments}.
In Section~\ref{sec:discuss}, we discuss the implications, limitations, and other possible extensions and applications of the estimator. 
\new{We summarize in Section~\ref{sec:summary}.}

\section{Motivation and Formulation} 
\label{sec:motiv}

In this \documentname, we use the following notation.
We write vectors in bold and lowercase, e.g. $\vv{}$; tensors in bold and uppercase, e.g. $\TT{}$; and unstructured data blobs in sans serif, e.g. $\GG{}$.
A hat above a symbol, e.g. $\bld{\hat{\xi}}$, indicates an estimate of the value.

\subsection{Standard Two-Point Correlation Function Estimation}
\label{sec:ls}

The standard approach to estimating the two-point correlation function involves counting pairs of tracers within a survey volume as a function of separation scale.
Let's assume we have a data catalog with $N_D$ objects within a sky volume.
We also require a random catalog with $N_R$ objects Poisson-distributed throughout the same volume.
We can define a set of separation bins which we will use to estimate the \cf at various scales.
We are then ready to sum in each bin the relevant pairs of objects within and across our catalogs.
In standard notation, these pair counts are written as $DD$, $DR$, and $RR$, as in \eqt{eq:lsintro}.
To clarify that these are in fact vectors, with length $K$ where $K$ is the number of bins, we use the symbol $\vv{}$; then, for example, the data--data pair counts $DD$ become $\vv{DD}$.
We can then write the \LS estimator as 
\begin{equation} \label{eq:ls}
    \bld{\hat{\xi}} = \frac{\vv{DD} - 2\,\vv{DR} + \vv{RR}}{\vv{RR}} ~.
\end{equation}
\new{We also make explicit the pair-count normalization factors, defining $\NN{DD} \equiv \frac{2}{\NN{D}\,(\NN{D}-1)}$, $\NN{DR} \equiv \frac{1}{\NN{D}\,\NN{R}}$, and $\NN{RR} \equiv \frac{2}{\NN{R}\,(\NN{R}-1)}$.}
The components of the pair-count vectors can then be written as
\begin{eqnarray}\displaystyle
    \label{eq:ls1}
    \left[ \vv{DD} \right]_k &\equiv& \frac{1}{\NN{DD}} \adjustlimits \sum_{n} \sum_{n' < n} i(g_k < |\bld{r}_n - \bld{r}_{n'}| < h_k) \\ 
    \left[ \vv{DR} \right]_k &\equiv& \frac{1}{\NN{DR}}\sum_{n} \sum_{m} i(g_k < |\bld{r}_n - \bld{r}_m| < h_k) \\
    \label{eq:ls3}
    \left[ \vv{RR} \right]_k &\equiv& \frac{1}{\NN{RR}} \adjustlimits\sum_{m} \sum_{m'<m} i(g_k < |\bld{r}_m - \bld{r}_{m'}| < h_k) ~,
\end{eqnarray}
where $\left[ \vv{} \right]_k$ is the pair counts in bin $k$ (which has bin edges $g_k$ and $h_k$), $i$ is an indicator function that returns $1$ if the the condition is true and otherwise returns $0$, $\bld{r}$ is the tracer position, the $n$ and $n'$ indices index data positions, and the $m$ and $m'$ indices index random catalog positions.
The sums are over unique pairs, and for auto-correlations they exclude self-pairs; the normalization prefactors then account for the total number of possible pairs, explaining the difference between the auto- and cross-correlation factors.
The tracer position can be in real or redshift space, or broken down into the transverse and line-of-sight directions in the anisotropic correlation function; in this \documentname we consider the isotropic real-space \cf for simplicity, but the estimators detailed here apply equally well to these alternative configurations.
The estimator is also easily applicable to cross-correlations of two datasets.
 
The \LS estimator is known to be optimal (i.e. it is unbiased and has minimum variance) under a particular set of conditions: in the limit of unclustered data, for a data volume much larger than the scales of interest, and an infinitely large random catalog. 
In practice the latter two limits are sufficiently satisfied, but the data we are interested in are clustered.
\cite{VargasMagana2013} show that for clustered data, the \LS estimator has lower variance than other estimators, but does not reach the Poisson noise limit.
When applied to clustered data, \LS does show a bias on very large scales ($>$130 \hmpc), but the bias is significantly smaller than that of most other estimators (\citealt{Kerscher1999}, \citealt{VargasMagana2013}).
\LS is also less sensitive to the number of random points than other estimators \citep{Kerscher2000}.
While \LS has been sufficient for past analyses, its persisting bias and suboptimal variance under imperfect conditions mean that improvement is possible, and will be necessary for realistic large-scale structure measurements on modern datasets.

\subsection{\Est}
\label{sec:est}

\new{We present an estimator, \est, that generalizes the \LS estimator in a manner inspired by least-squares fitting.
We first provide an intuitive description of the estimator, followed by the mathematical formulation; we then derive the connection to least-squares in Section~\ref{sec:leastsq}.}

\new{\Est replaces the binned pair counts of \LS with a \textit{projection} of the pairs onto any set of basis functions; the linear superposition of these projections is our estimate of the \cf.
Essentially, \est outputs the best-fit linear combination of basis functions to describe the \cf.
This is most straightforward to see for the case of tophat (rectangular) basis functions.
Each pair is projected onto the tophat functions, with one of the functions getting a contribution of 1 and the rest getting 0; the sum of these contributions for all pairs produces the total data projection onto each tophat.
Note that this is equivalent to slotting the pairs into separation bins and summing them to get the binned pair counts, as in standard binned estimators.
We can now replace the tophat functions with any set of basis functions; when we project the pairs onto these functions, the functions may get fractional contributions as they no longer have to be bin-like.
(For simplicity we consider basis functions that are only depend on pair separation, but in fact \est can perform projections onto bases that are functions of any pair properties.)
The sum of these contributions will give the total data projection onto the set of basis functions.}

\begin{figure}[t!]
    \centering
    \includegraphics[width=0.6\textwidth]{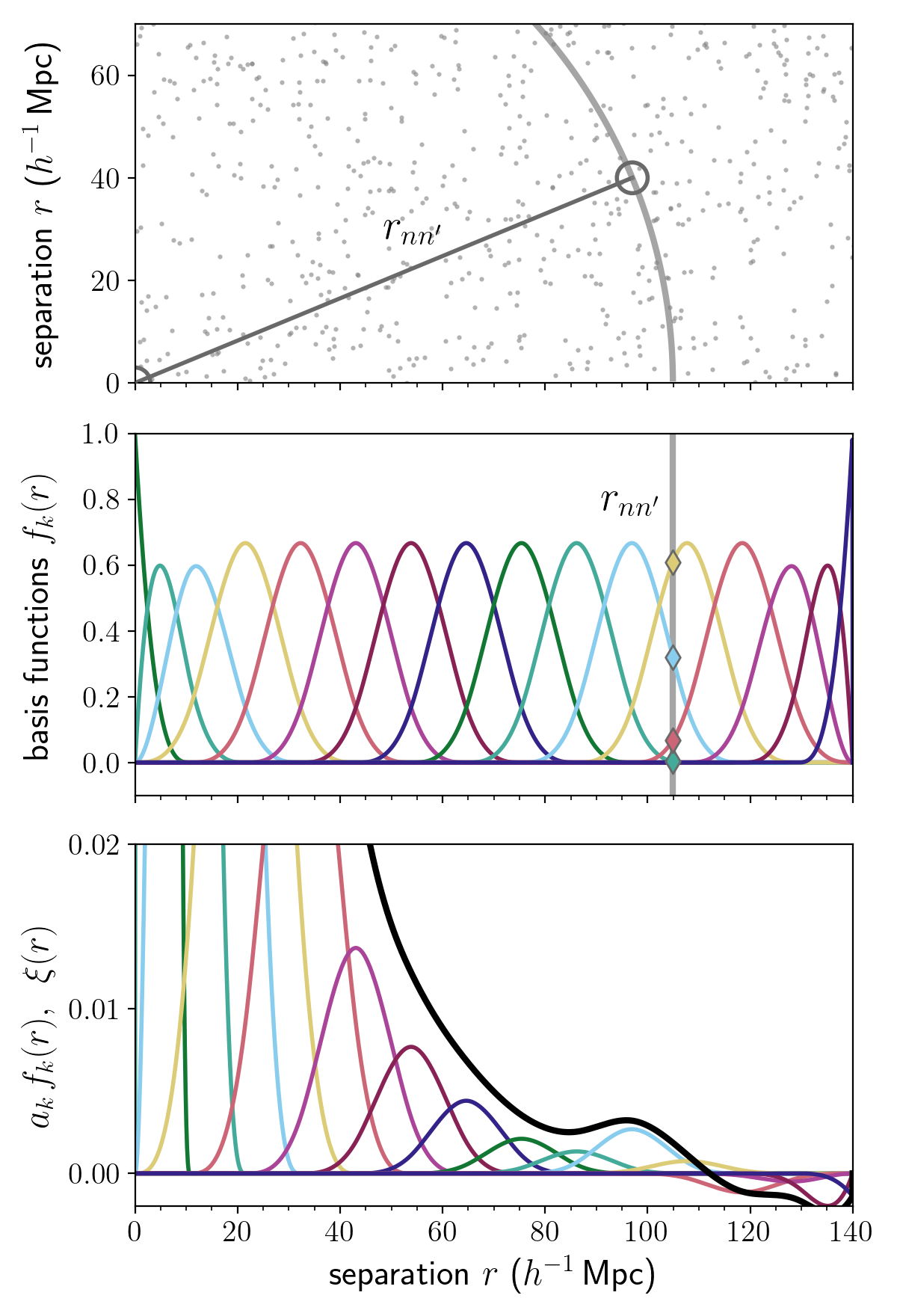}
    \caption{\new{A schematic illustrating how \est works. The top panel shows a slice of a dataset; each point represents a galaxy. One pair $\GG{n n'}$ is highlighted, with separation $r_{n n'}$ (grey line). The middle panel shows how \est performs the projection of the pair onto a chosen basis. Here we use cubic spline basis functions $f_k$ (see Section~\ref{sec:spline}; colored lines). The contribution of the pair to each basis function is given by the value of the function at separation $r_{n n'}$ (grey line); this quantity is $f_k(r_{n n'})$ (colored diamonds). The bottom panel shows the basis functions weighted by the final amplitudes $a_k$, given by the combination of contributions by all pairs. The final \cf estimate $\xi(r)$ (thick black line) is the sum of these, as given by \eqt{eq:xi_proj}.}}
    \label{fig:schematic}
\end{figure}

\new{We perform the same procedure with all pairs in the random catalog in order to obtain the random-random projection, and similarly with the data-random projection.
In general, the continuous functions will not be orthogonal or normalized, so we also require a normalization matrix.
This is constructed from the outer product of the random-random pair projections, for technical reasons described below.
We can then combine these projections in a manner similar to the pair-count combination of the \LS estimator, with the denominator replaced by the normalization matrix; this produces a final set of \textit{amplitudes} for the basis functions. 
The \cf estimate is simply the sum of the basis functions weighted by these amplitudes.
It can be evaluated at any pair separation, as the basis functions are continuous, producing a continuous correlation function. 
This process is shown schematically in Figure~\ref{fig:schematic}, with a choice of cubic spline basis functions (for more details on the spline see Section~\ref{sec:spline}).}

\new{We formalize this concept by generalizing the \LS estimator detailed in Section~\ref{sec:ls}.}
We take the indicator function $i$ to be any set of basis functions $\ff$, which returns a vector of length $K$, the number of bases.
We further generalize the arguments of the function to any properties of the galaxy pair, rather than just the separation between pairs; we call $\GG{ij}$ the data payload for the pair of galaxies $i$ and $j$.
This gives us, instead of pair counts, a vector $\vv{}$ of projections of the pairs onto basis functions, defined as
\begin{eqnarray}\displaystyle
    \label{eq:vdd}
    \vv{DD} &\equiv& \frac{1}{\NN{DD}} \adjustlimits \sum_{n} \sum_{n'<n} \ff(\GG{n n'}) \\
    \label{eq:vdr}
    \vv{DR} &\equiv& \frac{1}{\NN{DR}} \sum_{n} \sum_{m} \ff(\GG{n m}) \\
    \label{eq:vrr}
    \vv{RR} &\equiv& \frac{1}{\NN{RR}} \adjustlimits \sum_{m} \sum_{m'<m} \ff(\GG{m m'}) ~.
\end{eqnarray}
\new{We also require a normalization matrix to ensure that the estimator remains affine-invariant in the case of non-orthogonal basis functions.}
We thus define a projection tensor $\TT{RR}$ based on the autocorrelation of the random catalog, 
\begin{equation}
    \label{eq:Trr}
    \TT{RR} \equiv \frac{1}{\NN{RR}} \adjustlimits \sum_{m} \sum_{m'<m} \ff(\GG{m m'}) \cdot \ff\T(\GG{m m'}) ~,
\end{equation}
\new{where the exact formulation is motivated in Section~\ref{sec:leastsq} and satisfies the required affine invariance of the estimator as shown in Appendix~\ref{sec:affine}.}

We can now define \est as
\begin{equation}
    \label{eq:xi_proj}
    \hat{\xi}(\GG{\ell \ell'}) \equiv \bld{\hat a}\T \cdot \ff(\GG{\ell \ell'}) ~,
\end{equation}
where $\GG{\ell \ell'}$ contains the data values at which to evaluate $\bld{\hat{\xi}}$, and $\bld{\hat a}$ is a $K$-vector of the computed amplitudes of the basis functions
\begin{equation}
    \label{eq:amplitude}
    \bld{\hat a} \equiv \TT{RR}\inv \cdot (\vv{DD} - 2\,\vv{DR} + \vv{RR}) ~.
\end{equation}
\new{This parallels the pair-count combination of the \LS estimator, generalized to a least-squares fitting formulation.
As we will see in the next section,} the linear combination of vectors projects the pair information onto the features encoded by the basis functions, and the projection tensor performs a rescaling of these features.

We emphasize that $\GG{\ell \ell'}$ in \eqt{eq:xi_proj} does not represent a real pair of galaxies, but instead allows us to evaluate the \cf at any set of separations or other properties.
We can think of it as the data for an imaginary pair of galaxies $\ell$ and $\ell'$ that have a separation $r$ at which we want to evaluate $\bld{\hat{\xi}}$, and we would compute $\bld{\hat{\xi}}$ for such a pair at every separation in which we are interested.
As the formulation of \est is extremely general, we could choose basis functions that depend on other galaxy properties (see Section~\ref{sec:applications}); then, we would also choose each $\GG{\ell \ell'}$ pair to have values of these properties at which we want to evaluate $\bld{\hat{\xi}}$.
In the experiments in this \documentname, however, we will only take into account the separation between pairs, so we will write $\hat{\xi}(r)$.

\Est reduces to the \LS estimator when we choose $\ff$ to be a set of tophat functions in pair separation.
Explicitly, from our galaxy pair data $\GG{n n'}$, we use only their separation,  $|\bld{r}_n - \bld{r}_{n'}|$.
We can then define a set of $K$ basis functions $\ff$ as
\begin{equation}
    \label{eq:ff_separation}
    f_k(\GG{n n'}) =  i(g_k < |\bld{r}_n - \bld{r}_{n'}| < h_k) ~,
\end{equation}
where $k$ denotes the tophat component with edges $g_k$ and $h_k$ as before.
In this case the $\vv{DD}$, $\vv{DR}$ and $\vv{RR}$ projection vectors become binned pair counts.
The $\TT{RR}$ tensor becomes diagonal as the bins are orthogonal, with its diagonal elements equal to the elements of the $\vv{RR}$ vector.
\new{Then the amplitudes $\bld{\hat a}$ are exactly the \cf estimate given by \LS, and evaluating $\bld{\hat{\xi}}$ on a fine grid of pair separations results in a continuous step-function representation of the \LS estimate.}

\new{The fact that \est reduces to \LS motivates the understanding of our estimator as related to the limit of infinitesimal bins.
In the standard framework, it is impossible to work in this limit due to the extreme variance properties, but our approach resolves this by returning to the space of finite basis functions.
\Est effectively computes \LS in bins of infinitesimal width, and then rotates and stretches these bins into the space of a chosen set of basis functions.
The binned pair counts can then be re-summed into the form of the basis functions.
The estimator can perform such a rotation as it has the property of affine invariance, meaning it is invariant to rotations or stretches of the basis functions; we show this in Appendix~\ref{sec:affine}.
Thus, \est measures the projection of the data onto continuous basis functions by effectively working in the limit of infinitesimal bins and performing a rotation into the function space.} 

\Est can be straightforwardly generalized to cross-correlations between two datasets.
In this case, we consider datasets $D_1$ and $D_2$, and associated random catalogs $R_1$ and $R_2$. 
We then have cross-correlations rather than auto-correlations for the data-data and random-random terms, and two different data-random terms, crossing each dataset with the opposite random catalog. 
The data-data term becomes 
\begin{equation}
    \vv{D_1 D_2} \equiv \frac{1}{\NN{D_1}\,\NN{D_2}} \sum_{n_1} \sum_{n_2} \ff(\GG{n_1 n_2}) ~,
\end{equation}
where $n_1$ and $n_2$ index the data points in each catalog, and the normalization factor is now simply the inverse product of catalog sizes as we are no longer concerned with double-counting.
The other terms ($\vv{D_1 R_2}$, $\vv{D_2 R_1}$, $\vv{R_1 R_2}$, $\TT{R_1 R_2}$) generalize as one would expect.
The amplitudes then become
\begin{equation}\displaystyle
    \bld{\hat a} \equiv \TT{R_1 R_2}\inv \cdot (\vv{D_1 D_2} - \vv{D_1 R_2} - \vv{D_2 R_1} + \vv{R_1 R_2})
 \end{equation}
and we use this to compute the estimator as in \eqt{eq:xi_proj}.

Finally, we can write down the form of \est when we are working with a periodic box and the survey window is effectively infinite.
In this case, we can analytically compute the $\vv{DR}$, $\vv{RR}$, and $\TT{RR}$ terms.
The derivation and formulation of these terms are shown in Appendix~\ref{sec:analytic}.

We discuss the implementation of \est in Section~\ref{sec:comp}.

\subsection{Connection to Least-Squares Fitting}
\label{sec:leastsq}

\new{The formulation of \est is based on linear least-squares fitting; in this section we derive the connection between the two.
This derivation is closely connected to the derivation of a similar estimator given by \cite{Tessore2018}.}

\new{We can formulate the goal of clustering estimation as finding the best representation of spatial data in the space of two-point separation.
Recall that the linear least-squares fit to a set of data is
\begin{equation}
    \label{eq:leastsq}
    \bld{\hat{\theta}} = [\bld{X}^\mathsf{T}\,\bld{C}^{-1}\,\bld{X}]^{-1}\, [\mathbf{X}^\mathsf{T}\,\bld{C}^{-1}\,\bld{y}] ~,
\end{equation}
where $\bld{\hat{\theta}}$ is the vector of best-fit parameters, $\bld{X}$ is a design matrix containing functions of fitting features, $\bld{C}$ is the covariance matrix, and $\bld{y}$ is a column vector of data to be fit.
The second bracketed factor $[\bld{X}\T\,\bld{C}\inv\,\bld{y}]$ projects the data onto the features (as in a matched filter). 
The first bracketed factor $[\bld{X}\T\,\bld{C}\inv\,\bld{X}]\inv$ rescales the projected features into the space of the parameters.
Standard two-point correlation function estimators are effectively performing such a projection: each bin is the projection of the data pair counts onto the radial separation annulus, and the random-random term rescales this feature.
The analogy is clear in the so-called natural estimator of the \cf, $\bld{\hat{\xi}} = \vv{DD}/\vv{RR} - 1$ (e.g., \citealt{Kerscher2000}), with $\vv{DD}$ paralleling with the second bracketed factor and $\vv{RR}$ the first (the division can be written as an inverse factor).}

\new{We can make this parallel explicit by dividing our survey window into a fine grid of cells (or ``voxels'') such that each cell contains exactly 0 or 1 galaxies. 
Then we can consider the occupation number $\mathcal{N}$ of each cell, where $\mathcal{N}$ is 0 or 1.
We construct a design matrix $\bld{X}$ which contains the the features for each of the $\NN{CC}$ cell pairs, which has dimensions  $\NN{CC} \times K$, where $K$ is the number of features.
These features can be described as the values of basis functions $\ff$ given the properties $\GG{\ell \ell'}$ of the cell pair indexed by $\ell \ell'$ (properties that include, say, the separation $r$ between the cells),
\begin{equation}
    X_{(\ell \ell')k} = f_k(\GG{\ell \ell'}) ~,
\end{equation}
where we note that $\ell \ell'$ indexes a single row in $\bld{X}$, but contains two indices as it refers to a cell pair.
For a given galaxy catalog in this survey window, we can define a vector of observables $\bld{y}$ of length $\NN{CC}$.
Where the cell pair hosts a data pair, $y_{\ell \ell'} = 1$; elsewhere, $y_{\ell \ell'} = 0$.}

\new{With this notation, we can rewrite our projection vectors of equations (\ref{eq:vdd})--(\ref{eq:vrr}) as
\begin{eqnarray}
    \vv{DD} &\equiv& \frac{1}{\NN{DD}} \bld{X}\T\,\bld{y}_\mathrm{DD} \\
    \vv{DR} &\equiv& \frac{1}{\NN{DR}} \bld{X}\T\,\bld{y}_\mathrm{DR} \\
    \vv{RR} &\equiv& \frac{1}{\NN{RR}} \bld{X}\T\,\bld{y}_\mathrm{RR} ~.
\end{eqnarray}
We can define the projection tensor of \eqt{eq:Trr} as
\begin{equation}
    \TT{RR} \approx \frac{1}{\NN{CC}} \bld{X}\T\,\bld{X} ~.
\end{equation}
where the $\approx$ symbol indicates that we have taken the limit $\NN{RR}\to\NN{CC}$ in which the random catalog objects fill all cells in the grid.
We can now write our amplitudes of \eqt{eq:amplitude} as
\begin{equation}
    \label{eq:amps_lsq}
    \bld{\hat{a}} \approx \left[ \frac{1}{\NN{CC}} \, \bld{X}\T \, \bld{X} \right]\inv \left[ \frac{1}{\NN{DD}}\,\bld{X}\T\,\bld{y}_\mathrm{DD} 
    - 2\,\frac{1}{\NN{DR}}\,\bld{X}\T\,\bld{y}_\mathrm{DR} 
    + \frac{1}{\NN{RR}}\,\bld{X}\T\,\bld{y}_\mathrm{RR} \right] ~.
\end{equation}}

\new{The terms of \eqt{eq:amps_lsq} parallel the definition of generalized least-squares in \eqt{eq:leastsq}.
(Recall that we have taken the weights to be 1 for now, so our covariance matrix $\bld{C}$ here is just the identity matrix; non-uniform weights could straightforwardly be included with this formulation.)
Each term is the least-squares estimate for the \cf of the occupation number of the catalog pair.
For a given cell pair $\ell \ell'$, this quantity can be written $\langle \mathcal{N}_{\ell} \, \mathcal{N}_{\ell'} \rangle$, and we see that we can predict it at any cell pair by defining $X_{\ell \ell'}$ as the associated feature vector and taking the product of this with the amplitudes:
\begin{equation}
    \bld{X}_{\ell \ell'} \, \bld{\hat{a}} \, \simeq \, \frac{\NN{CC}}{\NN{DD}} \langle \mathcal{N}_{\mathrm{D},\ell} \, \mathcal{N}_{\mathrm{D},\ell'} \rangle 
    - 2\,\frac{\NN{CC}}{\NN{DR}}\,\langle \mathcal{N}_{\mathrm{D},\ell} \, \mathcal{N}_{\mathrm{R},\ell'} \rangle
    + \frac{\NN{CC}}{\NN{RR}} \langle \mathcal{N}_{\mathrm{R},\ell} \, \mathcal{N}_{\mathrm{R},\ell'} \rangle ~.
\end{equation}
Here we have used the $\simeq$ symbol to indicate that $\bld{X}_{\ell \ell'} \, \bld{\hat{a}}$ is an estimator for the quantity on the right-hand side.}

\new{We can understand the quantity $\langle \mathcal{N}_{\ell} \, \mathcal{N}_{\ell'} \rangle$ as the joint probability of finding galaxies in a pair of cells $\ell$ and $\ell'$.
For the random-random cross-correlation which has a vanishing correlation function, this is simply
\begin{equation}
    \langle \mathcal{N}_{\mathrm{R},\ell} \, \mathcal{N}_{\mathrm{R},\ell'} \rangle = \frac{\NN{RR}}{\NN{CC}} ~.
\end{equation}
We can make a similar statement for the data-random catalog which also has a vanishing correlation function.
For the data-data cross-correlation which has a non-vanishing correlation function $\xi_{\ell \ell'}$, this becomes (by the definition of \cf)
\begin{equation}
    \langle \mathcal{N}_{\mathrm{D},\ell} \, \mathcal{N}_{\mathrm{D},\ell'} \rangle = \frac{\NN{DD}}{\NN{CC}} \left( 1 + \xi_{\ell \ell'} \right) ~.
\end{equation}
Plugging these into the amplitude equation, we find
\begin{eqnarray}
    \bld{X}_{\ell \ell'} \, \bld{\hat{a}} \, &\simeq& \, \frac{\NN{CC}}{\NN{DD}} \left( \frac{\NN{DD}}{\NN{CC}} \left( 1 + \xi_{\ell \ell'} \right) \right)
    - 2\,\frac{\NN{CC}}{\NN{DR}}\,\left( \frac{\NN{DR}}{\NN{CC}} \right)
    + \frac{\NN{CC}}{\NN{RR}} \left( \frac{\NN{RR}}{\NN{CC}} \right) \\
    \label{eq:xi_lsq}
    \bld{X}_{\ell \ell'} \, \bld{\hat{a}} \, &\simeq& \, \xi_{\ell \ell'} ~.
\end{eqnarray}
We see that our amplitudes $\bld{\hat{a}}$ are the least-squares estimate of the correlation function, where we have assumed that we are working in the limit of a very large random catalog.
Equation (\ref{eq:xi_lsq}) shows that these best-fit amplitudes give the prediction of the correlation function $\hat{\xi}$ at a new cell pair, which we can choose to have any separation or other properties we like; this is precisely what we define in \eqt{eq:xi_proj} as the evaluation of the \cf.}

\section{Experiments and Results}
\label{sec:experiments}

\subsection{Lognormal Mock Catalogs}

We demonstrate the application of \est on a set of artificial data.
We generate lognormal mock catalogs \citep{ColesJones1991} using the \texttt{lognormal\_galaxies} code by \citep{Agrawal2017}.
We use an input power spectrum with the Planck cosmology, the same parameters used for the MultiDark--PATCHY simulations \citep{Kitaura2016} made for the Baryon Oscillation Spectroscopic Survey (BOSS, \citealt{Dawson2013}).
This assumes a cold dark matter model with $\Omega_m = 0.307115$, $\Omega_b = 0.048206$, $\sigma_8 = 0.8288$, $n_s = 0.9611$, and $h = 0.6777$.
Our fiducial test set is 1000 realizations of periodic cubes with size (750 \hmpc)$^3$ and a galaxy number density of $2 \times 10^{-4}$ $h^{3}\,$Mpc$^{-3}$.
We choose to perform these tests on periodic boxes so that we may compute the random--random term analytically (see Appendix~\ref{sec:analytic}), significantly cutting down on computation time.
The results will hold for catalogs with realistic survey windows and random--random terms computed directly with \est.

\subsection{Comparison of Standard Tophat Basis Functions}

\begin{figure}[ht]
    \centering
    \includegraphics[width=0.8\textwidth]{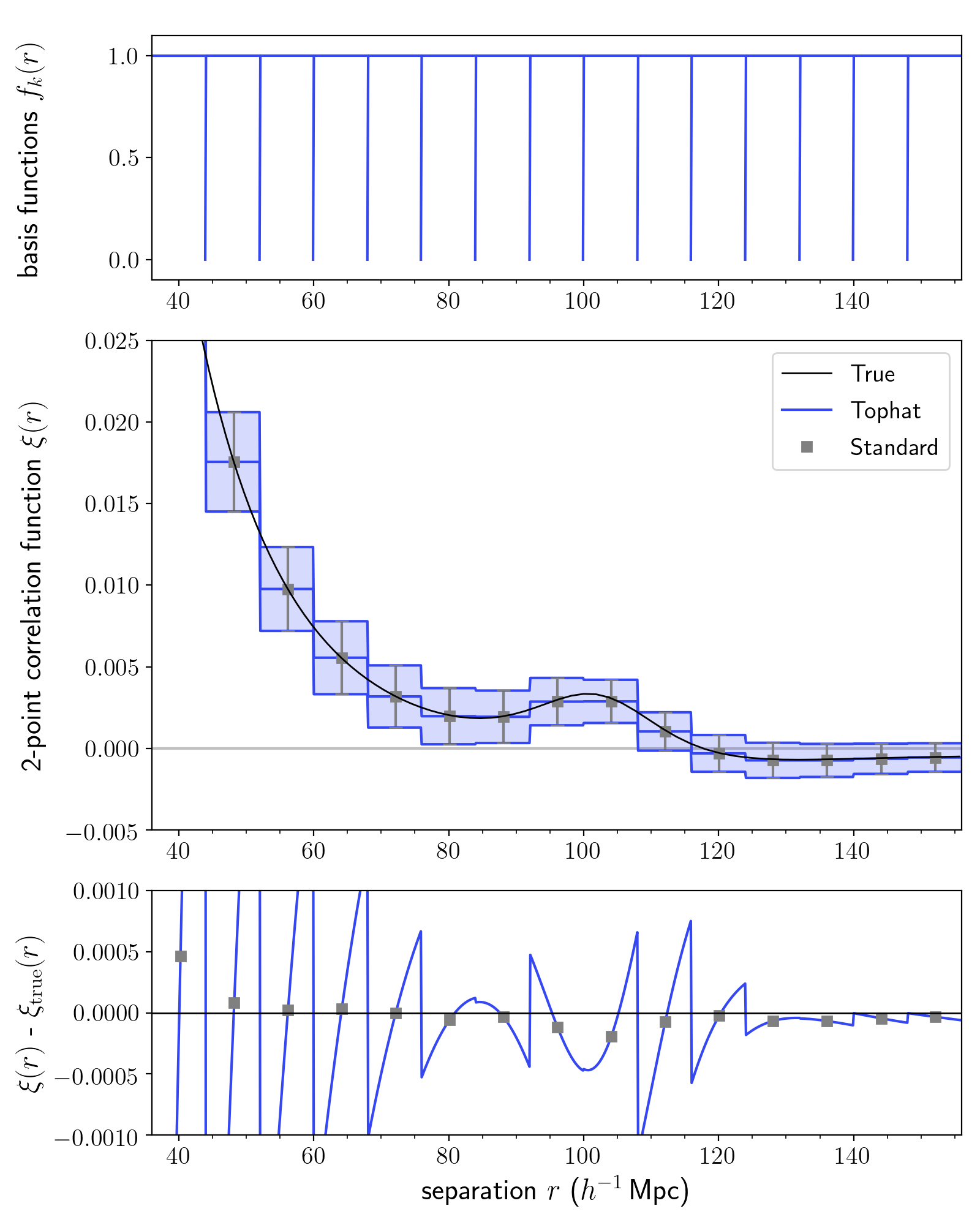}
    \caption{A comparison between \est with a tophat basis (thin blue lines) and the standard estimator (grey squares). The top panel shows the basis functions used for the tophat estimator. The middle panel shows the mean of the estimated correlation functions for 1000 mock catalogs, compared to the true input \cf (thin black line); the shaded region and error bars are the standard deviation of the \cf estimate. The lower panel shows the absolute error between the estimate and true \cf. \Est with a tophat basis is exactly equivalent to the standard estimator, but in a continuous form, emphasizing the fact that binning results in a poor representation of the true \cf.}
    \label{fig:tophat}
\end{figure}
    
We first estimate the correlation function of our mocks using the the standard estimator.
We choose 15 separation ($r$) bins in the range $36 < r < 156$ \hmpc, each with a width of 8 \hmpc; this was found to be the optimal bin width by \cite{Percival2014}, and is standard for two-point analyses.
We apply the estimator to each of our 1000 mock catalogs.
The mean of these estimated correlation functions is shown in Figure~\ref{fig:tophat}; the error bars show the standard deviation of the 1000 mocks in each bin.
We also show the true input correlation function, and the bottom panel shows the absolute error between the estimated and true correlation functions.

There remains an ambiguity in the $r$-value at which to plot the result of the standard estimator. 
The volume-weighted average is often used, or a weighted average depending on the pairs in the bin; this choice propagates to differences in comparing the estimate to models (though at the precision of current surveys these differences are not significant).
Here we plot the standard estimator with the volume-weighted average.

We demonstrate \est with a tophat basis function.
We choose tophat functions with the same locations and widths as the bins used for the standard estimator; these are shown in the top panel of Figure~\ref{fig:tophat}. 
As this estimator computes the \cf in a continuous form, we plot the result as a continuous function at every $r$-value.
In practice, this means choosing a fine grid of $r$-values at which to evaluate $\hat{\xi}(r)$; here we choose 1000 $r$-values across the separation range.
This results in a step function form for the correlation function.
The values of the \cf at each step exactly align with the result of the standard estimator.
In fact, this step function is exactly what the standard estimator is estimating; we have just made explicit the fact that the each estimate applies to the entire bin.
When we look at the error with respect to the truth (bottom panel), the error blows up at the edges of each bin, where the continuous estimate deviates most significantly from the truth.
\new{We recognize that in typical analyses one does not consider the error across the entire bin as shown; rather, one compares the \cf in the bin at some effective $r$-value to a model evaluated at that value.
However, we show the tophat errors here to emphasize  how the standard binned estimator is a poor representation of the true \cf.}

\subsection{Demonstration using Spline Basis Functions}
\label{sec:spline}

A natural extension of tophat basis functions is the B-spline.
B-splines of order $n$ are piecewise polynomials of order $n-1$; they constitute the basis functions for spline interpolation \citep{deBoor1987}.
They have the nice property that the functions and their derivatives can be continuous, depending on the order; \new{this might be important in inference contexts where it is useful to have differentiable models with respect to the parameters of interest, and generally aligns with our belief that physical models of the universe are analytic (i.e. have derivatives of all orders).}
Additionally, B-splines are well-localized, which provides a more direct comparison to the typical tophat basis (which is entirely localized).
For this demonstration we use fourth-order B-splines, which constitute the set of basis functions for a cubic spline, as they are the lowest-order spline to have a continuous first derivative.

\begin{figure}[ht]
    \centering
        \includegraphics[width=0.8\textwidth]{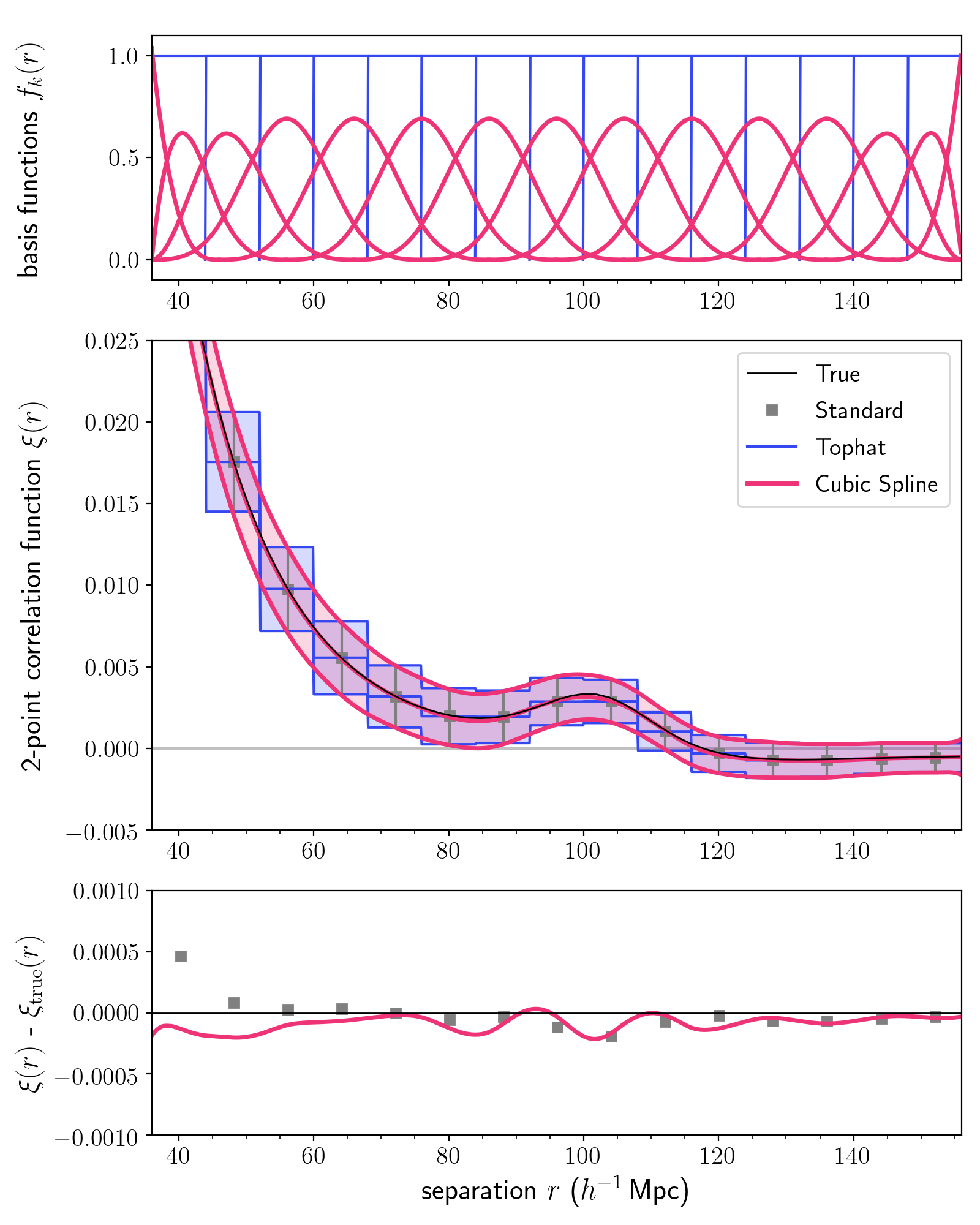}
        \caption{\new{A demonstration of} \est with cubic-spline basis functions (thick red). \new{We compare the cubic-spline estimate to the standard binned Landy--Szalay estimate (grey squares), and to the continuous-function estimate with a tophat function (thin blue).} The top panel shows the basis functions used for each measurement. The middle panel shows the mean of the estimated correlation functions for each of the 1000 mock catalogs compared to the true input \cf (thin black); the shaded region \new{and error bars} are the standard deviation. The lower panel shows the absolute error between the estimate and true \cf; \new{we do not show the tophat error for clarity}. It is clear that the spline basis function results in a correlation function that is a better representation of the true \cf in its shape and smoothness.}
        \label{fig:spline}
    \end{figure}

We compare the estimator with a cubic-spline basis to the \new{\LS estimator, both in the standard binned representation and reformulated as continuous functions using a tophat basis;} the results are shown in Figure~\ref{fig:spline}.
The basis functions are shown in the top panel of the figure.
We use the same tophat basis as above.
For the cubic-spline basis, we use the same $r$-range and number of basis functions, and knots chosen to evenly span the range. 
The cubic-spline bases on the edge have different shapes such that they remain normalized; we note that generally, one should choose the basis functions such that the \cf range of interest does not depend on the range of the basis functions.

\new{The middle panel shows that \est} using the cubic-spline basis clearly produces a better fit to the true correlation function in its shape and smoothness at every point across the scale range, compared to the estimator using the tophat basis.
The bottom panel shows the error with respect to the truth; the \new{cubic-spline estimator is similar in accuracy}, and directly comparable to the truth (or model) at every scale.
On the other hand, in order to compare the binned \cf to a model, one must integrate the model over the bin range, though in practice the model is often just evaluated at the effective $r$ of each bin.

This comparison demonstrates that there exist other sets of basis functions that produce better representations of the data compared to the standard tophat/binned estimator.
The choice of a high-order spline may be useful for cases in which one wants a mostly localized yet representative estimate of the \cf, or smooth derivatives.
Generally, the choice of basis functions should be tailored to the scientific goal; in the next section we explore the case of a BAO analysis.

\subsection{BAO Scale Estimation Test}
\label{sec:bao}

The measurement of the baryon acoustic oscillation (BAO) scale provides an apt use case for our estimator.
The BAO feature is a peak in clustering on large scales, $\sim$150 Mpc ($\sim$100\hmpc), making it less sensitive to small-scale astrophysical effects.
It is one of the best tools for constraining cosmological models, in particular the distance--redshift relation \citep{Cole2005, Eisenstein2005, Kazin2010, Anderson2012, Anderson2014, Alam2016}.

We base our BAO analysis on the method of the BOSS DR10 and 11 analysis \citep{Anderson2014}.
We estimate the spherically averaged 3-dimensional correlation function, $\hat{\xi}(r)$, where $r$ is the separation between pairs.
(BAO analyses are typically done in redshift space, estimating $\hat{\xi}(s)$, where $s$ is the redshift-space separation between pairs, but here we are using a periodic box in which we know the true galaxy positions, so we just use the real-space distance $r$.)
In order to extract information about the baryon acoustic feature from galaxy clustering, we must choose a fiducial cosmological model to convert redshifts to distances.
If we choose an incorrect model, the scales in the power spectrum will be dilated, so the oscillation wavelength---and thus the BAO peak position---will be shifted.
We can model this shift as a scale dilation parameter, $\alpha$, which is a function of the relevant distance scales in the true and fiducial cosmologies, defined as
\begin{equation} \label{eq:alpha}
\alpha = \Bigg( \frac{D_\mathrm{A}(z)}{D_\mathrm{A}^{\text{mod}}(z)} \Bigg)^{2/3} \Bigg( \frac{H^{\text{mod}}(z)}{H(z)} \Bigg)^{1/3} \Bigg( \frac{r_\mathrm{s}^{\text{mod}}}{r_\mathrm{s}} \Bigg) ~,
\end{equation}
where $D_\mathrm{A}$ is the angular diameter distance, $H$ is the Hubble parameter, $z$ is the redshift, $r_\mathrm{s}$ is the sound horizon scale at the drag epoch, and the superscript ``$\text{mod}$'' denotes the value for the chosen fiducial model (the non-superscripted parameters are the true values).
Qualitatively, if the fit prefers $\alpha>1$, this suggests the true position of the BAO peak is at a smaller scale than in the fiducial model, whereas if $\alpha<1$, the peak is at a larger scale.
With isotropic analyses, there is a degeneracy between $D_\mathrm{A}$ and $H$, so typically a combination of these values is reported; the degeneracy can be broken with anisotropic BAO analyses.
Our estimator could straightforwardly perform an estimate of the anisotropic correlation function, but for demonstration purposes we perform an isotropic analysis here and focus on the recovered value of $\alpha$.

In standard practice, the fitting function used to determine the value of $\alpha$ is 
\begin{equation}
\xi^{\mathrm{fit}}(r) = B^2 \xi^{\mathrm{mod}}(\alpha r) + \frac{a_1}{r^2} + \frac{a_2}{r} + a_3 ~,
\end{equation}
where $B$ is a constant that allows for a large-scale bias, and $a_1$, $a_2$, and $a_3$ are nuisance parameters to account for the broadband shape.
A $\chi^2$ fit is performed at intervals of $\Delta \alpha$, with $B$, $a_1$, $a_2$, and $a_3$ as free parameters. 
The resulting value for $\alpha$ is used to derive the actual values of the distance scales of interest.
\new{Typically, analyses have to correct for the broadening of the BAO peak due to nonlinear growth \citep{Eisenstein2007}; approaches include computing the model with a damping parameter, or performing density-field reconstruction before applying the estimator.
As we use lognormal mocks, we do not need to perform this step, but a realistic analysis must either used a damped model for the basis functions or reconstruct the density field.}

The form of the standard fitting function is well-suited to our estimator, as it is a few-parameter model with a linear combination of terms.
To use our estimator to estimate $\alpha$, we add a term that includes the partial derivative of the model with respect to $\alpha$.
This allows us to have fixed basis functions, and for an initial choice of $\alpha_\mathrm{guess}$, determine the change in this value needed to improve the fit. 
Our fitting function is then
\begin{equation} \label{eq:baoiter_fit}
\xi^\mathrm{fit}(r) = B^2\,\xi^\mathrm{mod}(\alpha_\mathrm{guess}\,r) + C\,k_0\,\frac{\dd \xi^\mathrm{mod}(\alpha_\mathrm{guess}\,r)}{\dd \alpha} + a_1\,\frac{k_1}{r^2} + a_2\,\frac{k_2}{r} + a_3\,k_3 ~,
\end{equation}
where $C$ is an additional coefficient that describes the contribution of the derivative term, and $k_0$, $k_1$, $k_2$, and $k_3$ are constants that determine the initial magnitude of the basis functions.
In this case, the free parameters are $B^2$, $C$, $a_1$, $a_2$, and $a_3$.
Note that in theory the choice of $k_i$ values shouldn't matter as the estimator is affine invariant (see Appendix~\ref{sec:affine}), but in practice reasonable choices are important for stability.
The adopted $k_i$ values are noted in Appendix~\ref{sec:baoiter}.

\begin{figure}[t]
    \centering
    \includegraphics[width=0.8\textwidth]{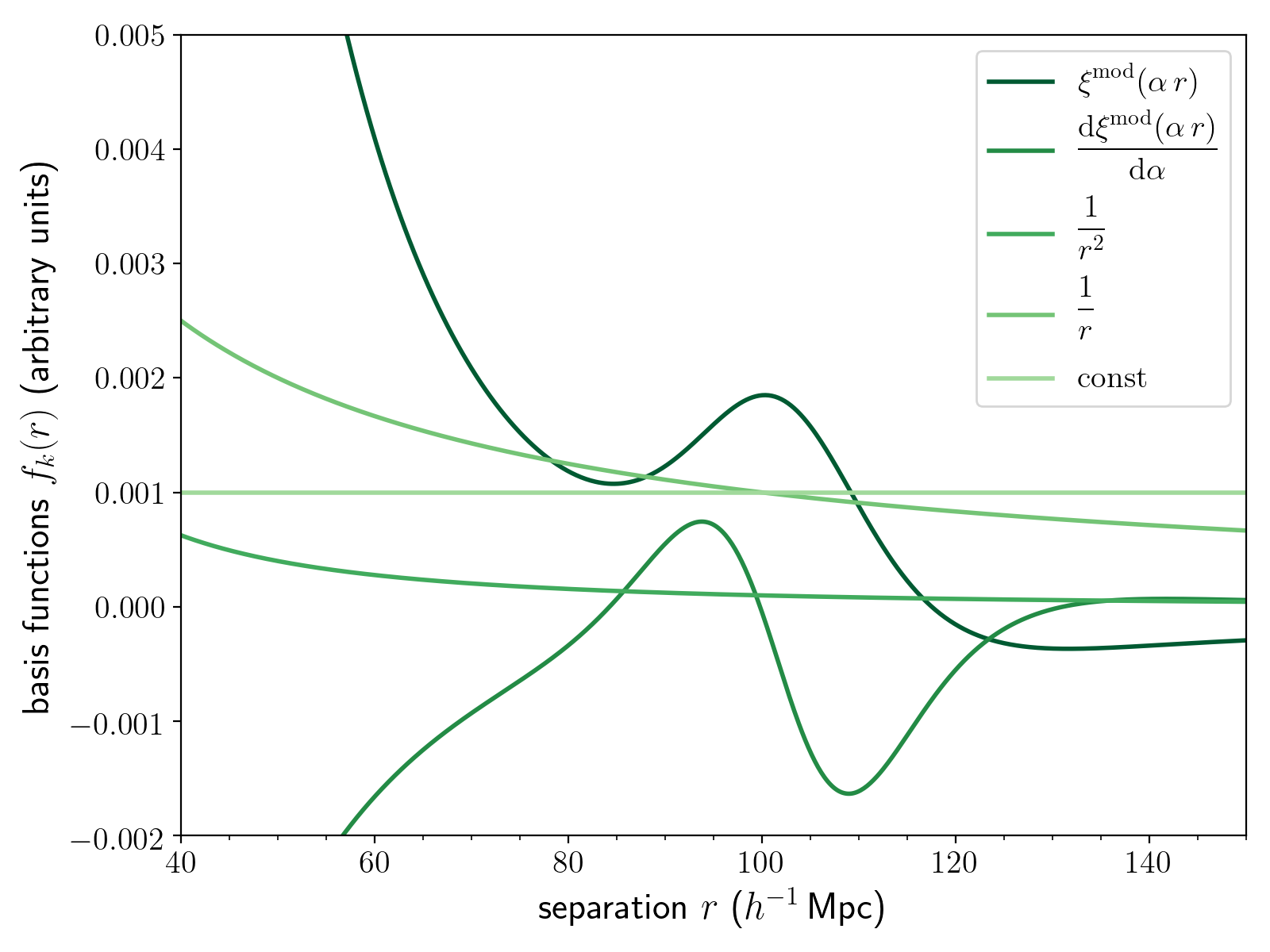}
    \caption{The set of basis functions used to fit for the BAO scale using our estimator. The $\xi^\mathrm{mod}(\alpha\,r)$ term (darkest green) is the correlation function computed using fiducial model, with some scale dilation $\alpha$. The derivative term (second-to-darkest green) is the derivative of this model with respect to $\alpha$, which allows for the direct estimation of this parameter. The other three terms (lighter greens) are nuisance parameters to fit the broadband shape.}
    \label{fig:bao_bases}
\end{figure}

To use the estimator for a BAO measurement, we input these five terms as the five basis functions of our estimator.
The estimator outputs an amplitude vector $\bld{\hat a}$ as described in Section~\ref{sec:est}, which describes the contribution of each basis function---precisely the values of the free parameters, scaled by $k_i$.
From the value of $C$, we can determine our estimate of the scale dilation parameter, $\hat{\alpha}$, as $\hat{\alpha} = \alpha_\mathrm{guess} + C\,k_0$, based on the definition of finite derivatives. 
With this formulation, a value of $C=0$ indicates that the current $\alpha_\mathrm{guess}$ gives the best fit to the data (given the chosen cosmological model), while nonzero values give the magnitude and direction of the necessary change in the scale dilation parameter to optimally fit the data.
In practice, we apply an iterative procedure to converge at our best estimate $\hat{\alpha}$; this procedure and other implementation details are described in Appendix~\ref{sec:baoiter}.

We demonstrate this method using our set of lognormal mock catalogs.
We construct a recovery test following that in \cite{Hinton2019}.
We assume the fiducial cosmological model used in \cite{Beutler2017}: $\Omega_{\text{m}} = 0.31$, $h = 0.676$, $\Omega_{\text{b}} = 0.04814$, $n_s = 0.97$. 
As we know the cosmology used for our mock catalogs, we can compute the true value of the scale dilation parameter, $\alpha_{\text{true}}=0.9987$.
(Here our choice of fiducial model happened to be close to the true model, so our $\alpha_{\text{true}}$ is very close to 1; this is typical, as our cosmological model is fairly well-constrained.)
With this fiducial model, we can construct the basis functions for our estimator; these are shown (with $\alpha=1$ and arbitrary scaling) in Figure~\ref{fig:bao_bases}.

\begin{figure}[t]
\centering
    \includegraphics[width=0.8\textwidth]{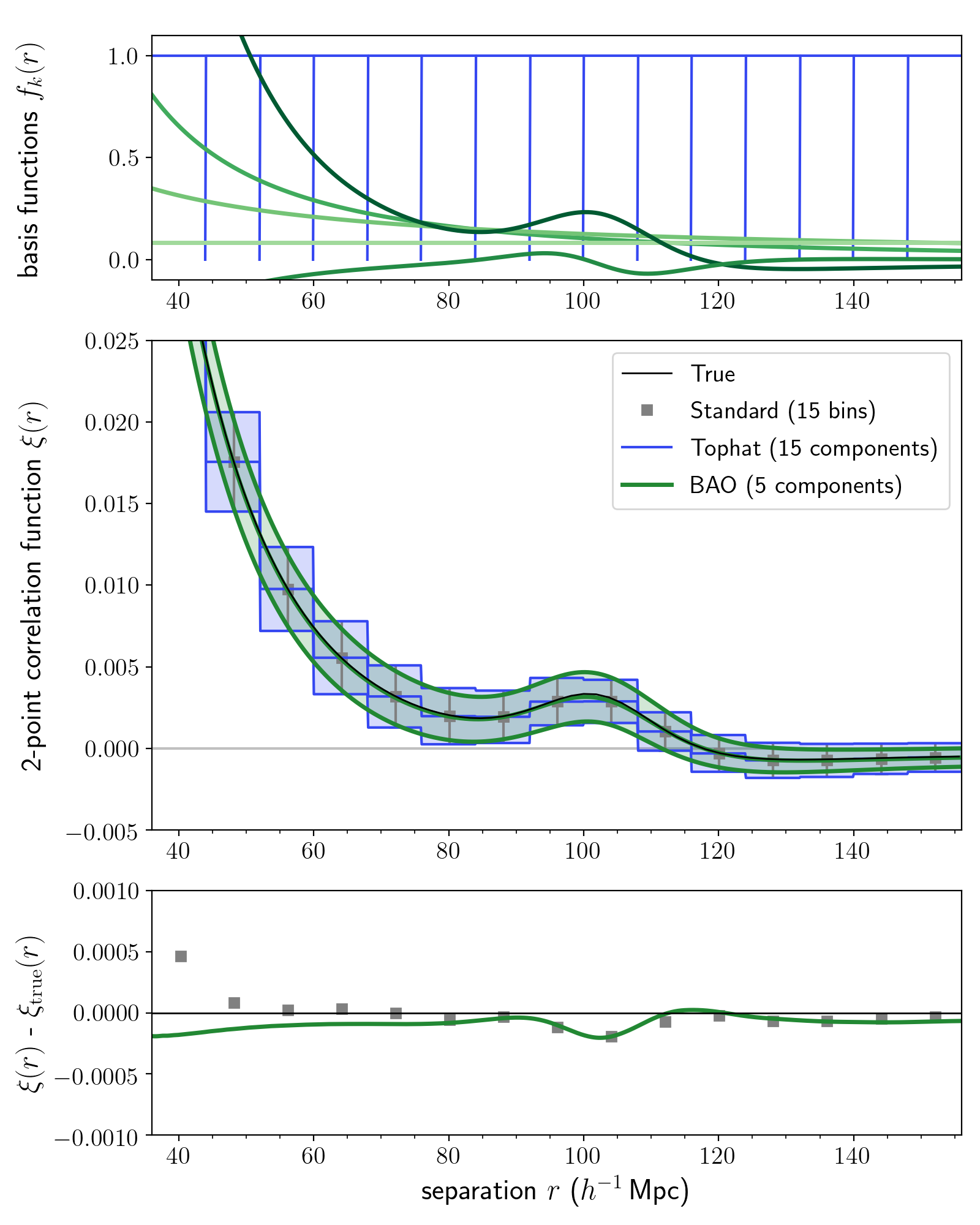}
    \caption{Estimation of the correlation function using \est with basis functions based on the BAO fitting function (thick green; same as in Figure~\ref{fig:bao_bases}, with arbitrary scaling). \new{We compare this to the standard binned Landy--Szalay estimator (grey squares), and to} \est with a tophat function (thin blue). \new{The panels are the same as in Figure~\ref{fig:spline}.}}
    \label{fig:bao}
\end{figure}

We apply our iterative estimation procedure to each of the 1000 mocks; the mean of the resulting estimates for the correlation function is shown in Figure~\ref{fig:bao}.
We show the BAO basis functions in the top panel, as in Figure~\ref{fig:bao_bases}.
We compare \est using the BAO bases with that using tophat bases, as in the previous sections; \new{we also show the standard representation of the binned estimator.}
The correlation function estimated with the BAO bases clearly produces a more representative estimate at all scales.
The estimate is also smoother than that produced using the cubic-spline basis functions (Figure~\ref{fig:spline}).
More importantly, it is scientifically motivated: the estimator produces the relative contributions of the terms of the BAO fitting functions.
\new{This gives us the best-fit model for each of the 1000 mock catalogs, without the need for intermediate binning and fitting steps, and} allows us to directly estimate the best-fit value of $\alpha$ for each one.
The median estimate of the recovered scale dilation parameter is $\alpha=0.9975 \pm 0.0290$, where the uncertainty is based on the 16th and 84th percentiles; our estimate is very close to the true value of $\alpha = 0.9987$.

We emphasize that the BAO-based estimate requires only five components, while the tophat basis requires 15 components (or bins, in the standard approach) in the same scale range.
This is critical for the efficient computation of a precise covariance matrix, as the errors depend on the number of components used for the estimate, as described in Section~\ref{sec:covariance}.
\Est with the BAO basis functions could reduce the number of mocks needed to achieve the same precision by a factor of a few to an order of magnitude; as these expensive cosmological simulations are currently the limiting step in two-point analyses, this could be highly impactful.

We note that these basis functions are significantly different than the tophat or B-spline bases previously explored.
One important difference is that they are not localized.
This means that data at all scales could contribute to all basis functions.
It is then critical for the investigator to ensure that the range of scales chosen is reasonable for the scientific problem at hand and that the final estimate of the parameter of interest does not depend on the details of this choice.
The nonlocality also means that the covariance matrix between the components will have a very different structure than typical binned covariance matrices.
We return to this issue in Section~\ref{sec:covariance}.

\section{Discussion} \label{sec:discuss}

\subsection{Estimator Usage and Limitations}

In this \documentname we have performed a few example applications of \est (for others see Section~\ref{sec:applications}); these demonstrate the problems solved and opportunities created with the estimator.
In removing the need for binning, the estimator can produce a much more \textit{representative} correlation function; a mixture of tophats is a very poor representation of the \cf. 
We almost always expect some sort of continuity in the functions we work with; nature does not bin.

One of the problems with binning is that there are choices to be made in bin widths and locations.
We note that our approach does not actually help with these choice issues.
In fact, \est expands investigator choice almost infinitely, as the space of possible basis functions is far larger than the space of possible binnings.
That said, \new{there are ways to make a well-motivated selection of basis functions.
In cases where there is a clear scientific model that can be represented as a linear combination of terms, choosing this will result in the most direct recovery of the parameters of interest (e.g., our BAO estimate in Section~\ref{sec:bao}).
When the investigator would like to remain more model-agnostic and obtain a continuous estimate, the choice of a basis such as splines, wavelets, Fourier components, or a power law may be appropriate.
There are various ways to compare the choice of basis functions; one approach is to perform cross-validation, which has proven useful in choosing histogram bins and kernels for density estimation \citep{Rudemo1982, Hogg2008}.}

\new{We also note that \est could be prone to both over- and under-fitting.
If a model that contains a particular feature is used directly as basis functions, and that feature is not present in the data, the estimator may output a \cf that is not representative of the data.
On the other hand, if a feature is not present in the basis functions but exists in the data, \est could smooth out or obscure it.
However, both of these issues arise in similar ways with the standard approach of binning and then fitting to a model that is not representative of the data, and binning may further erase features narrower than the bin width.
Still, it is true that the binned data may capture certain features that the chosen model doesn't.
Additional model flexibility could be included when using \est to address this concern.
We also suggest that investigators perform exploratory analysis with the standard estimator at various bin widths, and compare these to the results given by \est with the chosen set of basis functions.}

Our approach has a few other notable limitations.
One restriction is the fact that the \cf forms of interest must be representable in terms of a linear combination of basis functions; this makes \est less appropriate in the case of nonlinear models \new{(though often a linearization can be performed).}
\new{It is also difficult to restrict the space of the amplitudes, for instance to ensure a particular parameter remains positive, as could be done in binning-and-fitting approaches.}
The estimator must also evaluate the basis functions for every pair of objects, so highly complex functions may become intractable.
Finally, our estimator inherits many of the limitations of the Landy--Szalay estimator, including the fact that it contains a bias on clustered data and has non-optimal variance properties.
However, \est is further generalizable to other forms of the estimator (Section~\ref{sec:beyondls}), so in principle there is a formulation that further ameliorates these \LS-related limitations.

\subsection{Relationship to Other Estimators of Second-Order Statistics}
\label{sec:otherest}

On the surface, \est appears similar to \new{other second-order statistics} and two-point function projects.
\new{These include unbinned estimators for the correlation integral,} kernel density estimators, and the marked correlation function.
While \est shares some properties with these formulations, it ultimately solves a different problem.

\new{The problem of computing two-point statistics with non-trivial survey boundaries is a topic of significant study in the statistical point process literature \citep{Ripley1981, Illian2008, Diggle2013}.
Much of this work focuses on the correlation integral $C(r)$, which measures the average number of data points in a sphere of radius $r$ centered on a typical point; this is related by a factor of the density to the Ripley K-function \citep{Ripley1976}. These are closely connected to the two-point correlation function by an integral relation.
Estimators for $C(r)$ involve various ways of correcting for the survey boundary, such as local weights to correct for points near survey edges, and corrections based on the set covariance of the window function.
These typically do not require binning, as they are measuring an average of quantities local to each point, and can be computed at any value of $r$.
These $C(r)$ estimators are directly related to estimators for the \cf; the $DR$ and $RR$ pair counts are in fact Monte Carlo versions of the geometric $C(r)$ corrections mentioned \citep{Kerscher1999}.
However, the required shift to finite bins when estimating the two-point correlation function results in increased variance due to shot noise, and even the best estimator studied (\LS) is only unbiased in the limit of zero bin width for unclustered data.
As we rely the two-point correlation function for our standard cosmological analyses due to its theoretical interpretability, the unbinned correlation integral estimators do not represent viable alternatives to current \cf estimators.
While \est shares the avoidance of binning of these $C(r)$ estimators, it is not related in a more fundamental manner, instead generalizing directly from the the binned \LS estimator.}

Kernel density estimation (KDE) is a class of methods for estimating a probability density function from a set of data.
KDE methods essentially smooth the data with a given kernel, often a Gaussian.
This is useful when we want to reconstruct a distribution without making many assumptions about the data, as is required in parametric methods.
KDEs have found been applied in many areas of astrophysics, for example to measure the 21cm power spectrum with reduced foreground contamination \citep{Trott2019}, and to estimate luminosity functions with superior performance compared to binned methods \citep{Yuan2020}.
\cite{Hatfield2016} uses a KDE approach to estimate the angular correlation function, in order to address the issues of information loss and arbitrary bin choice inherent to binning; they optimize for the kernel choice, and find a correlation function consistent with that of the binned method.
\new{Another method closely related to KDE approaches involves the convolution of the 3-dimensional overdensity field with particular types of 3-dimensional kernels, even when only 2-dimensional transverse pair separations have been measured.
These have been used to make measurements of overdensities around individual galaxies \citep{Eisenstein2003}, amplitudes of small-scale clustering \citep{Padmanabhan2006}, and the scale of the baryon acoustic feature \citep{Xu2010}.}

\new{Some of these kernel-based methods are similar in spirit to \est, as they customize the functional form of the statistic to the scientific question of interest.}
However, KDE approaches perform a fundamentally different task: they smear out the data by taking the contribution of each data point to be a kernel function centered on that value, and sum these to determine the full distribution.
This achieves a smooth result, though in most contexts it produces biased estimates of the correlation function.
In contrast, \est projects each data point onto fixed basis functions, and uses the data to directly infer the best-fit contribution of each basis function.
This preserves the information in the data to the degree given by the chosen set of basis functions, which can in fact enhance features rather than smooth them.

Another method that shares similarities with \est is the marked correlation function (MCF, \citealt{Beisbart2000}; \citealt{Sheth2005}).
This estimator weights the two-point statistic by ``marks,'' which are typically properties of the tracers.
The MCF is useful for studying the connection between galaxies and their spatial clustering.
\cite{Skibba2006} used it to determine that luminosity-dependent clustering is a straightforward consequence of mass dependence.
\cite{Armijo2018} applied the MCF to test a class of modified gravity theories by marking with local density, demonstrating that there is additional information in the environmental dependence of clustering.
The MCF has also been shown to break the degeneracy between halo occupation distribution parameters and the cosmological parameter $\sigma_8$ \citep{WhitePadmanabhan2009}.

\Est can easily incorporate the \textit{idea} behind marks by choosing the basis functions to be functions of the desired properties of the tracer in addition to pair separation.
Combined with the choice of tophat basis functions and proper normalization, this would indeed be closely related to the measurements that can be made by the MCF.
However, \est can generalize this concept even further. 
Rather than still producing a two-point function that is only a function of separation, weighted by the marks, our estimator can elevate the marking properties to another continuous axis.
That is, it can estimate a multi-dimensional correlation function as a function of both separation and the given property.
This provides a more flexible way to look at the dependence of the \cf on the property, and has similar applicability to breaking parameter degeneracies.
We elaborate on the use cases for incorporating further tracer information into the choice of bases functions in Section~\ref{sec:applications}.

\subsection{Beyond the Landy--Szalay Estimator}
\label{sec:beyondls}

While we have formulated our estimator as a generalization of \LS, as this is the standard used in \cf analyses and has optimal properties under certain conditions, we can also reformulate it for other estimators.
Our formulation currently requires a normalization term (i.e. denominator) based on the random--random counts; for \LS we replace this with our $\TT{RR}$ term (\eqt{eq:Trr}).
This is also the case for the \cite{PeeblesHauser1974} (natural) estimator and the \cite{Hewett1982} estimator:
\begin{eqnarray}
    \bld{\hat{\xi}}_\text{P-H} &=& \frac{\vv{DD} - \vv{RR}}{\vv{RR}} \rightarrow \TT{RR}\inv \cdot \left( \vv{DD} - \vv{RR} \right)\\
    \bld{\hat{\xi}}_\text{Hew} &=& \frac{\vv{DD} - \vv{DR}}{\vv{RR}} \rightarrow \TT{RR}\inv \cdot \left( \vv{DD} - \vv{DR} \right) ~.
\end{eqnarray}
We can also straightforwardly generalize estimators which have a data--random cross-correlation as the normalization term, such as the \cite{DavisPeebles1983} estimator,
\begin{equation}
    \bld{\hat{\xi}}_\text{D-P} = \frac{\vv{DD} - \vv{DR}}{\vv{DR}} \rightarrow \TT{DR}\inv \cdot \left( \vv{DD} - \vv{DR} \right) ~,
\end{equation}
where we define
\begin{equation}
    \TT{DR} = \frac{1}{\NN{DR}} \sum_{n} \sum_{m} \ff(\GG{n m}) \cdot \ff\T(\GG{n m}) ~.
\end{equation}

The continuous form of these estimators can be extended to cross-correlations in the straightforward way expected.
This formulation could also be extended to nearly any linear combination of pair counts.
The estimator of \cite{VargasMagana2013}, for instance, selects the optimal combination of pair counts; our estimators could be combined to create an even more generalized estimator.
However, some estimator formulations use powers of these terms that are nontrivial to reformulate as normalization tensors for our estimation approach; we leave this problem for future work.

\subsection{Implementation and Computational Performance}
\label{sec:comp}

We implement \est within the correlation function package \texttt{Corrfunc} (\citealt{Sinha2019}, \url{https://github.com/manodeep/Corrfunc}).
\new{\texttt{Corrfunc} is the state-of-the-art package for computing correlation functions and other clustering statistics; it is written in C with python bindings and utilities.
It is used in many published analyses, and it is modular, user-friendly, and open-source.}
\texttt{Corrfunc} performs extremely fast pair identification, and can estimate correlation functions on both theoretical boxes and on-sky data.
Our implementation of \est capitalizes on the power of \texttt{Corrfunc}, adding the necessary functionality into the existing framework.
We allow the user to define the basis functions for continuous-function estimation, and for every pair identified, we pass the separation and any additional required tracer information to the basis functions.
We output the projection vector and the projection tensor, as well as the traditional pair counts.
We include additional functions to compute the amplitudes from these projections, and then to evaluate the continuous correlation function for these amplitudes at the desired points in parameter space, giving the user a high level of control at every step.
\new{Our implementation of \est is also open-source and available at \url{https://github.com/kstoreyf/suave}.}

The computational scaling for our estimator is by definition the same as the traditional method, as pair-finding remains the limiting factor.
However, because \est must evaluate the set of basis functions for each pair of galaxies, \new{it can decrease the computational performance (i.e. decrease the speed and increase the computational cost) compared to the traditional estimator}.
For simple basis functions like splines, this will only marginally decrease performance.
For more complicated functions, \est may incur significant extra computational expense.
Basis functions can also be input on a grid (of separation or any other property) and then interpolated; the performance is then similar for all functions, depending on how the interpolation is done, but interpolating each function for each pair does somewhat decrease the performance.
Though the performance at the time of estimation may be slower than the traditional estimator, the choice of basis may significantly save computational time in other areas, such as reducing the number of mock catalogs required for covariance matrix estimation; see Section~\ref{sec:covariance}.

We detail a number of implementation choices here.
Our formulation of \est requires the inverse of the random--random tensor $\TT{RR}$ to compute the amplitudes (\eqt{eq:amplitude}).
However, we don't compute this inverse directly, \new{as it can be unstable}, and is not in fact the end result we are interested in: we want the dot product between $\TT{RR}\inv$ and the numerator $\vv{}$ of the estimator.
For this reason, we use the ``solve'' operation which computes the solution $\bld{\hat a}$ of the well-determined matrix equation $\TT{RR}\,\bld{\hat a}=\bld{v}$.
We also make sure to report the condition number of the tensor, as numerical precision decreases as the condition number rises.
If the condition number large, then a rescaling of the basis functions or a rotation in the basis space can improve stability.

\subsection{Effect on Covariance Matrix Estimation}
\label{sec:covariance}

\new{We have shown that \est, with a proper choice of basis functions, results in \cf estimates that are just as accurate using fewer components compared to binned estimates (or, by extension, more accurate for the same number of components).}
This reduction in component number is critical when estimating the covariance matrix, which is required for standard parameter inference.
The covariance matrix is difficult to compute analytically, though there is promising progress on this front (e.g., \citealt{Wadekar2020}).
For major analyses, it is usually estimated by evaluating the \cf on a large number of mock catalogs and computing the covariance between the bins (e.g., \citealt{Reid2010}; \citealt{Anderson2014}).
The unbiased estimator for the sample covariance matrix is (e.g., \citealt{Anderson2003})
\begin{equation}
\big[ \bld{\hat{C}}^\mathrm{ML} \big]_{ij} = \frac{1}{\NN{mocks}-1} \sum_{q=1}^{\NN{mocks}} \bigg( \big[\bld{\xi}_q \big]_i - \bar{\bld{\xi}}_i \bigg) \bigg([\bld{\xi}_q \big]_j - \bar{\bld{\xi}}_j \bigg)\T ~,
\end{equation}
where $q$ denotes the index of the mock, $i$ and $j$ denote the index of the bin or component, $\bld{\xi}$ denotes the estimate in that bin for that mock, and $\bar{\bld{\xi}}$ denotes the mean value of the estimate in that bin across the mocks, where we have omitted the hat for clarity.

We typically require the inverse covariance matrix for analyses, but its form is nontrivial, as the inverse of an unbiased estimator is not necessarily unbiased.
Standard practice applies a correction factor \citep{Hartlap2007},
\begin{equation}
\bld{\hat{C}}\inv = \frac{\NN{mocks}-\NN{bins}-2}{\NN{mocks}-1} \left( \bld{\hat{C}}^\mathrm{ML} \right) \inv ~.
\end{equation}
However, this does not correct for errors in the covariance matrix; these propagate to the uncertainties on the estimated cosmological parameters, resulting in an overestimation of the error bars (\citealt{Dodelson2013}; \citealt{Percival2014}; \citealt{TaylorJoachimi2014}).
Assuming that $\NN{mocks} >> \NN{bins}$ (with both much larger than the number of parameters to be estimated), and that the measurements are Gaussian-distributed, the error bars are inflated by a factor of $(1 + \NN{bins}/\NN{mocks})$ (i.e., the true constraints are tighter than the derived ones).
This factor becomes critical at the precision of cosmological parameter estimation \citep{Percival2014}.

Typically, this is dealt with by generating a very large number of mocks.
For the Baryon Oscillation Spectroscopic Survey (BOSS, \citealt{Dawson2013}) DR9 analysis, 600 mocks were needed and the two-point correlation function used 41 bins \citep{Sanchez2012} (though they also perform a restricted 15-bin analysis over the BAO peak scales). 
For the BOSS DR14 fiducial \cf results, 1000 mocks and 18 bins were used \citep{Ata2017}.
Some surveys have already turned to approximate methods for these mock catalogs instead of performing full cosmological simulations, as the cost is prohibitive.
Future surveys will have even more costly requirements on mock catalogs, with larger simulations necessary to cover the larger survey volumes and more realistic mocks required to achieve the desired accuracy and precision on the covariance matrix.

An alternative to increasing $\NN{mocks}$ is decreasing $\NN{bins}$ to achieve the same error on precision.
In the binned standard method, this is shown to \emph{increase} the statistical variance, albeit only slightly \citep{Percival2014}.
A substantial increase in bin width would prevent capturing information in finer clustering features; even the relatively broad BAO peak requires a bin size on the order of its width of $\sim$10\hmpc.
In fact, in the standard method more bins would typically be desireable, but the number is limited by the available number of mocks for covariance matrix computation.

We have shown that we can use \est to estimate the \cf using fewer components, without sacrificing accuracy.
This means that we can safely reduce $\NN{bins}$, or in our case, the number of components (or basis functions) $K$.
The covariance matrix will then express the covariance between these components (rather than bins, as we have obviated binning).
To then achieve the same precision on the error on the cosmological parameters, a lower value of $\NN{mocks}$ becomes possible.
This will significantly reduce requirements on mock catalog construction, which will be particularly important for upcoming large surveys. 
Alternatively, with the same number of mock catalogs, one can achieve increased precision just using \est as an alternative to the standard estimator.
\new{We note that in addition to galaxy clustering analyses, this is particularly relevant to weak lensing surveys \citep{Mandelbaum2018a}, and the estimator presented here could be straightforwardly adapted to this application.}

Another issue with the covariance of the standard estimator is that the uncertainty is highly correlated across bins.
Thus the diagonal terms of the covariance matrix are poor representations of the true error on each bin.
The errors can be decorrelated by choosing a new estimator that is a linear combination of the original \cf bins. 
\cite{Hamilton2000} proposed a transformation using the symmetric square root of the Fisher matrix, and this was shown in \cite{Anderson2014} to significantly suppress the off-diagonal elements of the covariance matrix.
While this decorrelated covariance matrix is not used in the fitting in that analyses, it is useful for visualizing the uncertainty of the \cf estimates.
\Est could also be used to obtain a decorrelated covariance matrix.
One could perform an initial estimation with standard bins or basis functions, and then apply a transformation to decorrelation them.
These decorrelated bins could then be passed to the estimator as basis functions, and the analysis run again, in order to obtain a direct estimation that produces representative diagonal errors.
This might be particularly important for non-localized basis functions such as the BAO basis functions, which are expected to have highly correlated errors.
Any covariance estimate will have to estimate these covariances well, but of course this is also true for the tophat basis where the covariances are nontrivial.

\subsection{Further Applications}
\label{sec:applications}

The formulation of \est opens up many possibilities for extracting information from the correlation function.
The most straightforward applications are standard basis functions or linearizable astrophysical models, as we have shown for the standard BAO fitting function (Section~\ref{sec:bao}).

One natural set of applications is extensions of the isotropic real-space analyses presented in this work.
We could extend the spherically-averaged BAO analysis to a full anisotropic analysis, estimating the correlation function $\hat{\xi}(s, \mu)$ as a function of both the redshift-space separation $s$ and the cosine $\mu$ of the angle between $s$ and the line-of-sight direction (e.g. \citealt{Anderson2013}).
This is typically done with a fine binning along the $\mu$ direction and then compressed into angular moments via multipoles or clustering wedges \citep{Kazin2012}; \est could obviate this line-of-sight binning in addition to the binning along separation as we have shown.
Our estimator could similarly be extended to the 2-dimensional correlation function  $\hat{\xi}(r_p, \pi)$ used to separate out the effects of redshift-space distortions, with a parameterization of the pair separations parallel ($\pi$) and perpendicular ($r_p$) to the line of sight; this can then be integrated to give the projected real-space correlation function $w_p(r_p)$ \citep{DavisPeebles1983}.
This projected \cf provides another useful way to test $\Lambda$CDM (e.g. \citealt{Nuza2013}).
Our estimator could also be directly related to a power spectrum analysis by selecting a Fourier basis as our set of continuous functions, which would directly project the data onto Fourier modes.
This could represent a step towards unifying correlation function and power spectrum estimation.

\Est could be applied to the direct estimation of cosmological parameters beyond the distance-redshift relation.
For instance, one perform an analysis focused on the growth rate of cosmic structure $f$ \citep{Satpathy2016, Reid2018}, or the primordial non-Gaussianity in the local density field $f^{local}_{NL}$ \citep{Karagiannis2014}.
One could take this idea even further by choosing as basis functions a parametrized model of the \cf and the derivatives of this model with respect to all the parameters of interest, including cosmological parameters or halo occupation distribution parameters.
\Est would then directly output the projection of the \cf onto the derivative terms, which would then be translatable to changes in the  model that best fits the data.
This is analogous to the BAO analysis performed in Section~\ref{sec:bao}, but with a higher-dimensional space of derivatives of parameters of interest.
This approach would essentially perform a direct estimation of the parameters, without the need for the intermediate steps of binning and fitting.     

Another class of applications involves a choice of basis functions that depend not only on the separation between tracer pairs, but also on the properties of the tracers themselves.
One such use case is the redshift dependence of the Alcock--Paczynski effect \citep{AlcockPaczynski1979}, which can be used to constrain the matter density $\Omega_m$ and the dark energy equation of state parameter $w$ \citep{Li2016}.
The basis functions $f$ in this case would take the form
\begin{equation}
    \label{eq:ff_redshift}
    f_k(\GG{n n'}) = f_k(|\bld{r}_n - \bld{r}_{n'}|, z_n, z_{n'}) ~,
\end{equation}
where $z$ is the redshift of tracer $n$ or $n'$.
\Est would then output a \cf that is a function of both separation and redshift, providing a continuous way to look at the redshift dependence of clustering.

This approach would also lend itself to analyzing how the LSS relates to galaxy formation, a connection critical for understanding this astrophysical process.
The traditional way of doing this involves binning by galaxy luminosity, and then computing the correlation function of the galaxies in each bin (e.g., \citealt{Budavari2003}, \citealt{Zehavi2011}, \citealt{Durkalec2018}).
\Est can remove the need for this extra layer of binning by using basis functions that depend on both the pair separation and on some function of the luminosities of the two galaxies.
Similarly to the example of redshift dependence in \eqt{eq:ff_redshift} above, in this case the data payload $\GG{n n'}$ would contain the galaxy luminosities $L_n$ and $L_{n'}$ in addition to the pair separation $|\bld{r}_n - \bld{r}_{n'}|$, and the basis functions $f_k$ would take all of these as parameters.
This would result in a direct way to look at the \cf luminosity dependence.
This could be extended to other galaxy properties, such as color or Hubble type (e.g., \citealt{Li2006}, \citealt{Skibba2014}), as well as environmental properties like the local density (e.g., \citealt{Abbas2006}).
\Est provides the flexibility to explore such a high-dimensional parameter space, while binned methods become quickly limited by number statistics as one tries to include more parameters.

Beyond these standard use cases, the estimator gives us the opportunity to investigate more subtle or exotic signals which are anomalous with respect to our conventional models.
Anomalies could appear as inhomogeneities or anisotropies in the data.
For example, \cite{MukherjeeWandelt2018} investigated whether there is a directional dependence in estimated cosmological parameters across the sky, by performing analyses on patches of the Cosmic Microwave Background.
Another possibility is anisotropy in the cosmic acceleration, which could leave signatures in measurements made using various phenomena including baryon acoustic oscillations \citep{Faltenbacher2012} and Type Ia supernovae \citep{Colin2019}.
With our estimator, we could introduce a dependence on location or direction into our basis functions, and constrain the potential deviation from homogeneity or isotropy.
\Est would allow for a more precise estimate of this dependence as it doesn't require any sort of patches or spatial binning, instead estimating a multi-dimensional continuous \cf.
While these effects would be highly degenerate with systematics, our estimator combined with robust systematics mitigation opens investigation channels into new physics.

\section{Summary}
\label{sec:summary}

In this \documentname, we have presented a new approach to estimating the two-point correlation function for large-scale structure analyses, \est, without the need for binning.
It generalizes the standard \cf estimator in a manner inspired by least-squares fitting: it projects the data onto a set of user-chosen basis functions, and applies a normalization based on a random catalog.
\new{\Est has many advantages over traditional \cf estimation, including that it can
\begin{itemize}
    \item produce continuous two-point correlation functions that can be evaluated at any pair separation, removing the need for binning in  separation,
    \item incorporate other tracer properties into the basis functions, removing the need for binning along other axes (such as galaxy mass, or angle with respect to the line of sight),
    \item produce correlation functions that reflect our prior beliefs about the smoothness and shape of the \cf, given a well-motivated choice of basis functions,
    \item use basis functions that are tailored to the scientific goal, such as directly using the model as bases, avoiding an intermediate fitting step, and
    \item result in correlation functions that are more accurate with fewer components, lowering  requirements on costly mock catalog for covariance estimation.
\end{itemize}
There remain some limitations of \est, most notably that it inherits many of the issues of the Landy--Szalay estimator upon which it is based, including non-optimal variance properties, a bias at large scales, and a reliance on a random catalog.}

We demonstrated \est on a set of artificial mock catalogs.
We first showed that our method exactly reproduces the results of the standard approach with the choice of tophat basis functions, but in a way that demonstrates what the estimator is really measuring, namely a constant amplitude of clustering at every point within a radial separation bin.
We next showed that \est has the capacity to be much more expressive than the standard estimator: we demonstrated this with the choice of cubic-spline basis functions, which results in a correlation function estimate that is representative of the expected shape and smoothness of the \cf.
Further, we demonstrated that \est can be tailored to the scientific use case.
We applied to it a toy baryon acoustic feature analysis, choosing basis functions to be the terms of a modified BAO fitting function.
This produced an estimate of the \cf that inherently reflects our beliefs about its form, and resulted in a direct estimate of the scale dilation parameter with a high level of accuracy.

\new{\Est provides the opportunity for improved precision and accuracy in current and future large-scale structure surveys, as well as for the detection of novel signals in clustering analyses.}

\acknowledgements
KSF was supported by the Future Investigators in NASA Earth and Space Science and Technology (FINESST) award number 80NSSC20K1545 during the completion of this work.
KSF would like to acknowledge significant code feedback and support from Manodeep Sinha and Lehman Garrison.
The authors thank Michael Blanton, Jeremy Tinker, Roman Scoccimarro, David Grier, Alex Barnett, \new{Ashley Ross,} Lucia Perez, James Rhoads, Sangeeta Malhotra, Drew Jamieson, \new{Martin White, Nicholas Tessore, Ben Wibking}, Chris Lovell, and the members of the Flatiron Astronomical Data Group for helpful discussions and feedback.
\new{We also thank the anonymous referee for comments that improved the paper.}
All of the code used in this \documentname is available open-source at \url{https://github.com/kstoreyf/suave} and \url{https://github.com/kstoreyf/continuous-estimator}.

\appendix

\section{Affine Invariance of the Estimator}\label{sec:affine}

The estimate of the \cf with \est should not depend on the scaling of the chosen basis functions.
Thus we expect \est to be invariant under affine transformations of the basis functions, meaning transformations that preserve collinearity and distance ratios; the following demonstrates this affine invariance.

We represent the affine transformation by an invertible transformation matrix $\bld{M}$ that modifies the basis functions $\ff$, such that 
\begin{equation}
\ff' \leftarrow \bld{M}\,\ff ~,
\end{equation}
where the prime indicates our affine-transformed basis.
We choose $\bld{M}$ to be invertible to ensure the two bases have the same expressive capacity.
Then in the primed basis, the pair counts become
\begin{eqnarray}\displaystyle
\vv{DD}' &=& \adjustlimits \sum_{n} \sum_{n'} \ff_{n n'}' = \sum_{n n'} \bld{M}\,\ff_{n n'} = \bld{M}\,\vv{DD}
\\
\vv{DR}' &=& \sum_{n} \sum_{m} \ff_{n m}' = \sum_{n m} \bld{M}\,\ff_{n m} = \bld{M}\,\vv{DR}
\\
\vv{RR}' &=& \adjustlimits \sum_{m} \sum_{m'} \ff_{m m'}' = \sum_{m m'} \bld{M}\,\ff_{m m'} = \bld{M}\,\vv{RR} ~,
\end{eqnarray}
where we use the shorthand $\ff_{i j} = \ff(\GG{i j})$ and we have omitted the normalization factors for clarity.
In the last step, we have factored $\bld{M}$ out of the summation and written the primed projection vectors in terms of the unprimed vectors. 

For the random--random tensor we have
\begin{eqnarray}\displaystyle
\TT{RR}' &=& \adjustlimits \sum_{m} \sum_{m'} (\bld{M}\,\ff_{m m'}) \cdot (\bld{M}\,\ff_{m m'})\T \\
&=& \bld{M}\left[ \adjustlimits \sum_{m} \sum_{m'} \ff_{m m'} \cdot \ff_{m m'}\T \right] \bld{M}\T \\
&=& \bld{M}\,\TT{RR}\,\bld{M}\T ~.
\end{eqnarray}
Then the amplitudes in the primed basis become
\begin{eqnarray}\displaystyle
\bld{\hat a}' &=& \TT{RR}\invp \cdot (\vv{DD}' - 2\,\vv{DR}' + \vv{RR}') \\
\bld{\hat a}' &=& [\bld{M} \TT{RR} \bld{M}\T]\inv \cdot [\bld{M}\,\vv{DD} - 2\,\bld{M}\,\vv{DR} + \bld{M}\,\vv{RR}] \\
&=& (\bld{M}\T)\inv \, \TT{RR}\inv \, \bld{M}\inv \cdot \bld{M}\,[\vv{DD} - 2\,\vv{DR} + \vv{RR}] \\
&=& (\bld{M}\T)\inv \, \TT{RR}\inv \cdot [\vv{DD} - 2\,\vv{DR} + \vv{RR}] \\
&=& (\bld{M}\T)\inv \, \bld{\hat a}
\end{eqnarray}
and the estimator $\bld{\hat{\xi}}'$ in the primed basis, using the shorthand $\hat{\xi}_{\ell \ell'} = \hat{\xi}(\GG{\ell \ell'})$, is 
\begin{eqnarray}\displaystyle
\hat{\xi}_{\ell \ell'}' &=& \bld{\hat a}\Tp \cdot \ff_{\ell \ell'} \\
\hat{\xi}_{\ell \ell'}' &=& [(\bld{M}\T)\inv \, \bld{\hat a}]\T \cdot (\bld{M}\,\ff_{\ell \ell'}) \\
&=& \bld{\hat a}\T \, [(M\inv)\T]\T \cdot (\bld{M}\,\ff_{\ell \ell'}) \\
&=& \bld{\hat a}\T \, \bld{M}\inv \cdot \bld{M}\,\ff_{\ell \ell'} \\
&=& \bld{\hat a}\T \cdot \ff_{\ell \ell'} \\
&=& \hat{\xi}_{\ell \ell'} ~.
\end{eqnarray}
Thus after an affine transformation of the basis function, the resulting estimator is equivalent to the estimator in the original basis.
The method is shown to be affine invariant.

\section{Computing the Random--Random Terms Analytically}\label{sec:analytic}

The autocorrelation of the random catalog is meant to approximate the autocorrelation of the window function. 
When we have a periodic cube, we can compute this $\vv{RR}$ term analytically in the standard approach to correlation function estimation.
Here we derive this, and then derive the equivalent for our continuous-basis $\vv{RR}$ and $\TT{RR}$ terms.

Our goal is to estimate the normalized number of pairs in a periodic cubic volume filled uniformly with tracers, $\vv{RR}^\mathrm{ana}$. 
We first consider an annulus indexed by $k$ around a single galaxy, with radial edges $g_k$ and $h_k$. 
This annulus has a volume $V_k$.
Taking the box to have an average number density $\bar{n}$, the number of galaxies expected in the annulus is $N_k = V_k \, \bar{n}$, and thus our selected galaxy contributes $N_k$ pairs to the count.   
We do this for each of the $\NN{R}-1$ other galaxies, and after including a factor of $\frac{1}{2}$ to account for the fact that this double counts pairs, we find a total pair count of $\frac{1}{2} \, (\NN{R}-1) \, N_k = \frac{1}{2} \, (\NN{R}-1) \, V_k \, \bar{n}$.
For a cubic volume, $\bar{n} = \NN{R}/L^3$, so our final pair count for the annulus is  $\frac{1}{2} \, \NN{R}(\NN{R}-1) \, V_k \, / L^3$.
We want the normalized pair counts, so we divide by the number of unique pairs, $\NN{R}(\NN{R}-1)/2$; this cancels out all instances of $\NN{R}$, as it should as this approximation is not actually using random catalog, and leaves us with $\left[ \vv{RR}^\mathrm{ana} \right]_k = V_k / L^3 ~.$

We now need to compute $V_k$; for hard-edged radial bins, we can compute $V_k$ simply as the difference between spherical volumes. 
We can represent this more generally as an integral,
\begin{equation} \label{eq:vol_tophat}
V_k = \int_{g_k}^{h_k} dV = 4\pi \int_{g_k}^{h_k} r^2 \, dr ~,
\end{equation}
where we assume spherical symmetry.
We can generalize this to any basis function $f_k(r)$ that is only a function of $r$,
\begin{equation}
V_k = 4\pi  \int_{g_k}^{h_k} f_k(r) \, r^2 \, dr ~,
\end{equation}
where $k$ is now the index of the basis functions.
This reduces to \eqt{eq:vol_tophat} when $f_k(r)$ is the tophat function (returning 1 or 0 depending on whether or not $r$ falls between $g_k$ and $h_k$).

Combining the above equations gives us our full generalized analytic random--random projection vector $\vv{RR}^\mathrm{ana}$, which has elements
\begin{equation}
\left[ \vv{RR}^\mathrm{ana} \right]_k = \frac{1}{L^3} \, 4\pi \, \int_{r_\mathrm{min}}^{r_\mathrm{max}} f_k(r) \, r^2 \, dr ~,
\end{equation}
where we are now integrating over all values of $r$ we are interested in from some $r_\mathrm{min}$ to $r_\mathrm{max}$.
(For non-localized basis functions, the fully correct thing would be to integrate from $-\infty$ to $\infty$, though some bounds must be chosen in practice.)

Based on the definition of $\TT{RR}$ in \eqt{eq:Trr} as the sum of outer products of the basis function vectors and their transposes, the elements of the analytic random--random tensor $\TT{RR}^\mathrm{ana}$ can be written as
\begin{equation}
\left[ \TT{RR}^\mathrm{ana} \right]_{kk'} = \frac{1}{L^3} \, 4\pi \, \int_{r_\mathrm{min}}^{r_\mathrm{max}} f_k(r) \, f_{k'}(r) \, r^2 \, dr ~.
\end{equation}
This could be further generalized to account for basis functions that take properties other than pair separation as input.

For the periodic box case, we can think about the cross-correlation term $\vv{DR}$ similarly; now we are taking each of the $\NN{D}$ galaxies in the data catalog and estimating the normalized pair counts of their cross-correlation with a theoretical random catalog.
It turns out that the resulting projection vector for the cross-correlation is exactly the same as for the random autocorrelation, $\vv{RR}^\mathrm{ana}  = \vv{DR}^\mathrm{ana}$.
Thus the Landy-Szalay estimator reduces to the natural estimator, and we can compute the analytic amplitudes $\bld{\hat a}^{\mathrm{ana}}$ for \est as
\begin{equation}
\bld{\hat a}^{\mathrm{ana}} = \left[ \TT{RR}^\mathrm{ana} \right]\inv \cdot \left( \vv{DD} - \vv{RR}^\mathrm{ana} \right) ~.
\end{equation}
Finally, we use these amplitudes to compute the correlation function $\bld{\hat{\xi}}^{\mathrm{ana}}$ as before in \eqt{eq:xi_proj}.

This analytic form for the continuous estimator could be extended to basis functions that depend on other tracer properties in addition to pair separation.
In this case, one would have to integrate over these axes as well, but the idea is the same.

\section{Implementation of Estimation with BAO Basis Functions}\label{sec:baoiter}

\subsection{Iterative Procedure}

\Est can be used to measure the baryon acoustic oscillation (BAO) scale by choosing the basis functions to terms of a BAO fitting function, as described in Section~\ref{sec:bao} and shown in Figures~\ref{fig:bao_bases} and~\ref{fig:bao}.
For this application, we need to choose a fiducial cosmology for our bases, which will be offset from the true cosmology.
This offset can be encoded by a scale dilation parameter $\alpha$, which contains the information about the BAO scale; see \eqt{eq:alpha}. 
As our fitting function requires a fiducial model and an initial guess of this parameter, $\alpha_\mathrm{guess}$, and then determines the change needed, an iterative procedure is needed to converge to the best-fit value.

We start with assuming that we have chosen our fiducial model to match our true cosmology (we in all likelihood have not, but it's not a bad initial guess), giving us an initial $\alpha_\mathrm{guess} = 1.0$. 
We then apply \est to perform the measurement, and obtain the magnitude of the projection $C$ for the derivative term in our model as in \eqt{eq:baoiter_fit}. 
This gives us our estimate $\hat{\alpha}$ of the scale dilation parameter from this initial model; for the $i$th iteration, we have
\begin{equation}
    \hat{\alpha}_{i} = \alpha_{\mathrm{guess},i} + C_i \, k_0 ~,
\end{equation}
where $k_0$ is the chosen scaling parameter for the derivative basis function as in \eqt{eq:baoiter_fit}.

We choose the convergence criterion to be when the fractional change in $\hat{\alpha}$ between subsequent iterations falls below a threshold, $c_\mathrm{thresh}$,
\begin{equation}
    \left| \frac{\hat{\alpha}_i - \hat{\alpha}_{i-1}}{\hat{\alpha}_i} \right| < c_\mathrm{thresh} ~.
\end{equation}
For our application we choose $c_\mathrm{thresh} = 0.00001$.

To achieve convergence, we need to be careful in choosing our next $\alpha_{\mathrm{guess},i}$.
If it is far from the best estimate, $C_i$ will be large, and our resulting estimate $\hat{\alpha}_{i}$ will be inaccurate.
We thus include a damping parameter $\eta$ between 0 and 1 to improve our convergence.
Our next guess is then
\begin{equation}
    \alpha_{\mathrm{guess},i+1} \leftarrow \alpha_{\mathrm{guess},i} + \eta\,C_i\,k_0 ~.
\end{equation}
The choice of $\eta$ is important for stability and speed of convergence; too large a value can lead to a back-and-forth cycle in which the result hops between two values and never converges, and too small a value would make convergence take a very long time.
In our application, we start with $\eta=0.5$.
We check if our estimate is jumping over the true value by checking if the error changes sign; if it does, we reduce $\eta$ by a factor of $0.75$.

\subsection{Implementation Details}

We implement the partial derivative in the fitting function of \eqt{eq:baoiter_fit} as a finite difference between model with the our chosen value of $\alpha_\mathrm{guess}$, and the model with a value shifted by a small $\Delta \alpha$,
\begin{equation}
    \frac{\dd \xi^\mathrm{mod}(\alpha \, r)}{\dd \alpha} \leftarrow \frac{\xi^\mathrm{mod}(\alpha_\mathrm{guess} \, r) - \xi^\mathrm{mod}((\alpha_\mathrm{guess} + \Delta \alpha) \, r)}{\Delta \alpha} ~.
\end{equation}
In our implementation we take $\Delta \alpha = 0.001$; we have checked that our results are insensitive to this choice.

We choose the magnitudes of the basis functions $k$ to set them at similar scales, providing improved stability.
We use the values $k_0=0.1$, $k_1=10.0$, $k_2=0.1$, and $k_3=0.001$, though we check that the results are insensitive to choices near these values.

\bibliography{LSS-est_paper}

\begin{thebibliography}{}
\expandafter\ifx\csname natexlab\endcsname\relax\def\natexlab#1{#1}\fi
\providecommand{\url}[1]{\href{#1}{#1}}
\providecommand{\dodoi}[1]{doi:~\href{http://doi.org/#1}{\nolinkurl{#1}}}
\providecommand{\doeprint}[1]{\href{http://ascl.net/#1}{\nolinkurl{http://ascl.net/#1}}}
\providecommand{\doarXiv}[1]{\href{https://arxiv.org/abs/#1}{\nolinkurl{https://arxiv.org/abs/#1}}}

\bibitem[{Abbas \& Sheth(2006)}]{Abbas2006}
Abbas, U., \& Sheth, R.~K. 2006, Monthly Notices of the Royal Astronomical
  Society, 372, 1749, \dodoi{10.1111/j.1365-2966.2006.10987.x}

\bibitem[{Aghamousa {et~al.}(2016)Aghamousa, Aguilar, Ahlen, Alam, Allen,
  {Allende Prieto}, Annis, Bailey, Balland, Ballester, Baltay, Beaufore, Bebek,
  Beers, Bell, {Luis Bernal}, Besuner, Beutler, Blake, Bleuler, Blomqvist,
  Blum, Bolton, Briceno, Brooks, Brownstein, Buckley-Geer, Burden, Burtin,
  Busca, Cahn, Cai, Cardiel-Sas, Carlberg, Carton, Casas, Castander,
  Cervantes-Cota, Claybaugh, Close, Coker, Cole, Comparat, Cooper, Cousinou,
  Crocce, Cuby, Cunningham, Davis, Dawson, de~la Macorra, {De Vicente},
  Delubac, Derwent, Dey, Dhungana, Ding, Doel, Duan, Ealet, Edelstein,
  Eftekharzadeh, Eisenstein, Elliott, Escoffier, Evatt, Fagrelius, Fan,
  Fanning, Farahi, Farihi, Favole, Feng, Fernandez, Findlay, Finkbeiner,
  Fitzpatrick, Flaugher, Flender, Font-Ribera, Forero-Romero, Fosalba, Frenk,
  Fumagalli, Gaensicke, Gallo, Garcia-Bellido, Gaztanaga, {Pietro Gentile
  Fusillo}, Gerard, Gershkovich, Giannantonio, Gillet, Gonzalez-de Rivera,
  Gonzalez-Perez, Gott, Graur, Gutierrez, Guy, Habib, Heetderks, Heetderks,
  Heitmann, Hellwing, Herrera, Ho, Holland, Honscheid, Huff, Hutchinson,
  Huterer, {Seong Hwang}, {Maria Illa Laguna}, Ishikawa, Jacobs, Jeffrey,
  Jelinsky, Jennings, Jiang, Jimenez, Johnson, Joyce, Jullo, Juneau, Kama,
  Karcher, Karkar, Kehoe, Kennamer, Kent, Kilbinger, Kim, Kirkby, Kisner,
  Kitanidis, Kneib, Koposov, Kovacs, Koyama, Kremin, Kron, Kronig,
  Kueter-Young, Lacey, Lafever, Lahav, Lambert, Lampton, Landriau, Lang, Lauer,
  {Le Goff}, {Le Guillou}, {Le Van Suu}, {Hyeon Lee}, Lee, Leitner, Lesser,
  Levi, Li, Liang, Lin, Linder, Loebman, Luki, Ma, MacCrann, Magneville,
  Makarem, Manera, Manser, Marshall, Martini, Massey, Matheson, McCauley,
  McDonald, McGreer, Meisner, Metcalfe, Miller, Miquel, Moustakas, Myers, Naik,
  Newman, Nichol, Nicola, {Nicolati da Costa}, Nie, Niz, Norberg, Nord, Norman,
  Nugent, Oh, {G Olsen}, Padilla, Padmanabhan, Padmanabhan,
  Palanque-Delabrouille, Palmese, Pappalardo, Pris, Park, Patej, Peacock,
  Peiris, Peng, Percival, Perruchot, Pieri, Pogge, Pollack, Poppett, Prada,
  Prakash, Probst, Raichoor, {Hee Ree}, Refregier, Regal, Reid, Reil, Rezaie,
  Rockosi, Roe, Ronayette, Roodman, Ross, Ross, Rossi, Rozo, Ruhlmann-Kleider,
  Rykoff, Sabiu, Samushia, Sanchez, Sanchez, Schlegel, Schneider, Schubnell,
  Secroun, Seljak, Seo, Serrano, Shafieloo, Shan, Sharples, Sholl, Shourt,
  Silber, Silva, Sirk, Slosar, Smith, Smoot, Som, Song, Sprayberry, Staten,
  Stefanik, Tarle, {Sien Tie}, Tinker, Tojeiro, Valdes, Valenzuela, Valluri,
  Vargas-Magana, Verde, Walker, Wang, Wang, Weaver, Weaverdyck, Wechsler,
  Weinberg, White, Yang, Yeche, Zhang, Zhao, Zheng, Zhou, Zhou, Zhu, Zou, \&
  Zu}]{Aghamousa2016}
Aghamousa, A., Aguilar, J., Ahlen, S., {et~al.} 2016.
\newblock \doarXiv{1611.00036v2}

\bibitem[{Agrawal {et~al.}(2017)Agrawal, Makiya, Chiang, Jeong, Saito, \&
  Komatsu}]{Agrawal2017}
Agrawal, A., Makiya, R., Chiang, C.~T., {et~al.} 2017, Journal of Cosmology and
  Astroparticle Physics, 2017, \dodoi{10.1088/1475-7516/2017/10/003}

\bibitem[{Alam {et~al.}(2016)Alam, Ata, Bailey, Beutler, Bizyaev, Blazek,
  Bolton, Brownstein, Burden, Chuang, Comparat, Cuesta, Dawson, Eisenstein,
  Escoffier, Gil-Mar{\'{i}}n, Grieb, Hand, Ho, Kinemuchi, Kirkby, Kitaura,
  Malanushenko, Malanushenko, Maraston, Mcbride, Nichol, Olmstead, Oravetz,
  Padmanabhan, Palanque-Delabrouille, Pan, Pellejero-Ibanez, Percival,
  Petitjean, Prada, Price-Whelan, Reid, Rodr{\'{i}}guez-Torres, Roe, Ross,
  Ross, Rossi, Rubi{\~{n}}o-Mart{\'{i}}n, S{\'{a}}nchez, Saito,
  Salazar-Albornoz, Samushia, Satpathy, Sc{\'{o}}ccola, Schlegel, Schneider,
  Seo, Simmons, Slosar, Strauss, Swanson, Thomas, Tinker, Tojeiro, {Vargas
  Maga{\~{n}}a}, Al, \& Zhao}]{Alam2016}
Alam, S., Ata, M., Bailey, S., {et~al.} 2016, {The clustering of galaxies in
  the completed SDSS-III Baryon Oscillation Spectroscopic Survey: cosmological
  analysis of the DR12 galaxy sample}, Tech. rep.
\newblock \doarXiv{1607.03155v1}

\bibitem[{Alcock \& Paczynski(1979)}]{AlcockPaczynski1979}
Alcock, C., \& Paczynski, B. 1979, {An evolution free test for non-zero
  cosmological constant}, Tech. rep.

\bibitem[{Anderson {et~al.}(2011)Anderson, Aubourg, Bailey, Bizyaev, Blanton,
  Bolton, Brinkmann, Brownstein, Burden, Cuesta, {A da Costa}, Dawson,
  de~Putter, Eisenstein, Gunn, Guo, Hamilton, Harding, Ho, Honscheid, Kazin,
  Kirkby, Kneib, Labatie, Loomis, Lupton, Malanushenko, Malanushenko,
  Mandelbaum, Manera, Maraston, McBride, Mehta, Mena, Montesano, Muna, Nichol,
  Nuza, Olmstead, Oravetz, Padmanabhan, Palanque-Delabrouille, Pan, Parejko,
  P{\^{a}}ris, Percival, Petitjean, Prada, Reid, Roe, Ross, Ross, Samushia,
  S{\'{a}}nchez, Schlegel, Schneider, Sc{\'{o}}ccola, Seo, Sheldon, Simmons,
  Skibba, Strauss, {C Swanson}, Thomas, Tinker, Tojeiro, {Var-gas
  Maga{\~{n}}a}, Verde, Wagner, Wake, Weaver, Weinberg, White, Xu, Zehavi, \&
  Zhao}]{Anderson2012}
Anderson, L., Aubourg, E., Bailey, S., {et~al.} 2011, {The clustering of
  galaxies in the SDSS-III Baryon Oscillation Spectroscopic Survey: Baryon
  Acoustic Oscillations in the Data Release 9 Spectroscopic Galaxy Sample},
  Tech. rep.
\newblock \doarXiv{1203.6594v1}

\bibitem[{Anderson {et~al.}(2013)Anderson, Aubourg, Bailey, Beutler, Bolton,
  Brinkmann, Brownstein, Chuang, Cuesta, Dawson, Eisenstein, Honscheid, Kazin,
  Kirkby, Manera, Mcbride, Mena, Nichol, Olmstead, Padmanabhan,
  Palanque-Delabrouille, Percival, Prada, Ross, Ross, S{\'{a}}nchez, Samushia,
  Schlegel, Schneider, Seo, Strauss, Thomas, Tinker, Tojeiro, Verde, Weinberg,
  Xu, \& Eche}]{Anderson2013}
---. 2013, {The clustering of galaxies in the SDSS-III Baryon Oscillation
  Spectroscopic Survey: Measuring D A and H at z = 0.57 from the Baryon
  Acoustic Peak in the Data Release 9 Spectroscopic Galaxy Sample}, Tech. Rep.
  0000.
\newblock \doarXiv{1303.4666v1}

\bibitem[{Anderson {et~al.}(2014)Anderson, Aubourg, Bailey, Beutler, Bhardwaj,
  Blanton, Bolton, Brinkmann, Brownstein, Burden, Chuang, Cuesta, Dawson,
  Eisenstein, Escoffier, Gunn, Guo, Ho, Honscheid, Howlett, Kirkby, Lupton,
  Manera, Maraston, McBride, Mena, Montesano, Nichol, Nuza, Olmstead,
  Padmanabhan, Palanque-Delabrouille, Parejko, Percival, Petitjean, Prada,
  Price-Whelan, Reid, Roe, Ross, Ross, Sabiu, Saito, Samushia, S{\'{a}}nchez,
  Schlegel, Schneider, Scoccola, Seo, Skibba, Strauss, Molly, Thomas, Tinker,
  Tojeiro, Maga{\~{n}}a, Verde, Wake, Weaver, Weinberg, White, Xu, Y{\`{e}}che,
  Zehavi, \& Zhao}]{Anderson2014}
Anderson, L., Aubourg, {\'{E}}., Bailey, S., {et~al.} 2014, Monthly Notices of
  the Royal Astronomical Society, 441, 24, \dodoi{10.1093/mnras/stu523}

\bibitem[{Anderson(2003)}]{Anderson2003}
Anderson, T. 2003, {An Introduction to Multivariate Statistical Analysis},
  \dodoi{10.1080/00401706.1986.10488123}

\bibitem[{Armijo {et~al.}(2018)Armijo, Cai, Padilla, Li, \&
  Peacock}]{Armijo2018}
Armijo, J., Cai, Y.~C., Padilla, N., Li, B., \& Peacock, J.~A. 2018, Monthly
  Notices of the Royal Astronomical Society, 478, 3627,
  \dodoi{10.1093/MNRAS/STY1335}

\bibitem[{Ata {et~al.}(2017)Ata, Baumgarten, Bautista, Beutler, Bizyaev,
  Blanton, Blazek, Bolton, Brinkmann, Brownstein, Burtin, Chuang, Comparat,
  Dawson, de~la Macorra, Du, {du Mas des Bourboux}, Eisenstein,
  Gil-Mar{\'{i}}n, Grabowski, Kneib, Laurent, {Le Goff}, McEwen, Mueller,
  Myers, Newman, Tinker, Tojeiro, {Vargas Maga{\~{n}}a}, Vivek, Wang, Eche, Yu,
  Zarrouk, Zhao, Zhao, \& Zhu}]{Ata2017}
Ata, M., Baumgarten, F., Bautista, J., {et~al.} 2017, MNRAS, 000, 2.
\newblock \doarXiv{arXiv:1705.06373v2}

\bibitem[{Bailoni {et~al.}(2016)Bailoni, {Spurio Mancini}, Amendola, Bailoni,
  {Spurio Mancini}, \& Amendola}]{Bailoni2016}
Bailoni, A., {Spurio Mancini}, A., Amendola, L., {et~al.} 2016, {Improving
  Fisher matrix forecasts for galaxy surveys: window function, bin
  cross-correlation, and bin redshift uncertainty}, Tech. rep.
\newblock \doarXiv{1608.00458v3}

\bibitem[{Baxter \& Rozo(2013)}]{BaxterRozo2013}
Baxter, E.~J., \& Rozo, E. 2013, Astrophysical Journal, 779, 15,
  \dodoi{10.1088/0004-637X/779/1/62}

\bibitem[{Beisbart \& Kerscher(2000)}]{Beisbart2000}
Beisbart, C., \& Kerscher, M. 2000, The Astrophysical Journal, 545, 6,
  \dodoi{10.1086/317788}

\bibitem[{Beutler {et~al.}(2017)Beutler, Seo, Saito, Chuang, Cuesta,
  Eisenstein, Gil-Mar{\'{i}}n, Grieb, Hand, Kitaura, Modi, Nichol, Olmstead,
  Percival, Prada, S{\'{a}}nchez, Rodriguez-Torres, Ross, Ross, Schneider,
  Tinker, Tojeiro, \& Vargas-Maga{\~{n}}a}]{Beutler2017}
Beutler, F., Seo, H.~J., Saito, S., {et~al.} 2017, Monthly Notices of the Royal
  Astronomical Society, 466, 2242, \dodoi{10.1093/mnras/stw3298}

\bibitem[{Brout {et~al.}(2020)Brout, Hinton, \& Scolnic}]{Brout2020}
Brout, D., Hinton, S., \& Scolnic, D. 2020.
\newblock \doarXiv{2012.05900}

\bibitem[{Budavari {et~al.}(2003)Budavari, Connolly, Szalay, Szapudi, Csabai,
  Scranton, Bahcall, Brinkmann, Eisenstein, Frieman, Fukugita, Gunn, Johnston,
  Kent, Loveday, Lupton, Tegmark, Thakar, Yanny, York, \&
  Zehavi}]{Budavari2003}
Budavari, T., Connolly, A.~J., Szalay, A.~S., {et~al.} 2003, The Astrophysical
  Journal, 595, 59, \dodoi{10.1086/377168}

\bibitem[{Cole {et~al.}(2005)Cole, Percival, Peacock, Norberg, Baugh, Frenk,
  Baldry, Bland-Hawthorn, Bridges, Cannon, Colless, Collins, Couch, Cross,
  Dalton, Eke, {De Propris}, Driver, Efstathiou, Ellis, Glazebrook, Jackson,
  Jenkins, Lahav, Lewis, Lumsden, Maddox, Madgwick, Peterson, Sutherland, \&
  Taylor}]{Cole2005}
Cole, S., Percival, W.~J., Peacock, J.~A., {et~al.} 2005, Monthly Notices of
  the Royal Astronomical Society, 362, 505,
  \dodoi{10.1111/j.1365-2966.2005.09318.x}

\bibitem[{Coles \& Jones(1991)}]{ColesJones1991}
Coles, P., \& Jones, B. 1991, Monthly Notices of the Royal Astronomical
  Society, 248, 1, \dodoi{10.1093/mnras/248.1.1}

\bibitem[{Colin {et~al.}(2019)Colin, Mohayaee, Rameez, \& Sarkar}]{Colin2019}
Colin, J., Mohayaee, R., Rameez, M., \& Sarkar, S. 2019, Astronomy and
  Astrophysics, 631, \dodoi{10.1051/0004-6361/201936373}

\bibitem[{Colless {et~al.}(2001)Colless, Dalton, Maddox, Sutherland, Norberg,
  Cole, Bland-Hawthorn, Bridges, Cannon, Collins, Couch, Cross, Deeley, {De
  Propris}, Driver, Efstathiou, Ellis, Frenk, Glazebrook, Jackson, Lahav,
  Lewis, Lumsden, Madgwick, Peacock, Peterson, Price, Seaborne, \&
  Taylor}]{Colless2001}
Colless, M., Dalton, G., Maddox, S., {et~al.} 2001, Monthly Notices of the
  Royal Astronomical Society, 328, 1039,
  \dodoi{10.1046/j.1365-8711.2001.04902.x}

\bibitem[{Davis \& Peebles(1983)}]{DavisPeebles1983}
Davis, M., \& Peebles, P. J.~E. 1983, The Astrophysical Journal Supplement
  Series, 267, 465, \dodoi{10.1086/190860}

\bibitem[{Dawson {et~al.}(2013)Dawson, Schlegel, Ahn, Anderson, Aubourg,
  Bailey, Barkhouser, Bautista, Beifiori, Berlind, Bhardwaj, Bizyaev, Blake,
  Blanton, Blomqvist, Bolton, Borde, Bovy, Brandt, Brewington, Brinkmann,
  Brown, Brownstein, Bundy, Busca, Carithers, Carnero, Carr, Chen, Comparat,
  Connolly, Cope, Croft, Cuesta, Dacosta, Davenport, Delubac, {De Putter},
  Dhital, Ealet, Ebelke, Eisenstein, Escoffier, Fan, {Filiz Ak}, Finley,
  Font-Ribera, G{\'{e}}nova-Santos, Gunn, Guo, Haggard, Hall, Hamilton, Harris,
  Harris, Ho, Hogg, Holder, Honscheid, Huehnerhoff, Jordan, Jordan, Kauffmann,
  Kazin, Kirkby, Klaene, Kneib, Legoff, Lee, Long, Loomis, Lundgren, Lupton,
  Maia, Makler, Malanushenko, Malanushenko, Mandelbaum, Manera, Maraston,
  Margala, Masters, McBride, McDonald, McGreer, McMahon, Mena,
  Miralda-Escud{\'{e}}, Montero-Dorta, Montesano, Muna, Myers, Naugle, Nichol,
  Noterdaeme, Nuza, Olmstead, Oravetz, Oravetz, Owen, Padmanabhan,
  Palanque-Delabrouille, Pan, Parejko, P{\^{a}}ris, Percival,
  P{\'{e}}rez-Fournon, P{\'{e}}rez-R{\`{a}}fols, Petitjean, Pfaffenberger,
  Pforr, Pieri, Prada, Price-Whelan, Raddick, Rebolo, Rich, Richards, Rockosi,
  Roe, Ross, Ross, Rossi, Rubi{\~{n}}o-Martin, Samushia, S{\'{a}}nchez, Sayres,
  Schmidt, Schneider, Sc{\'{o}}ccola, Seo, Shelden, Sheldon, Shen, Shu, Slosar,
  Smee, Snedden, Stauffer, Steele, Strauss, Streblyanska, Suzuki, Swanson, Tal,
  Tanaka, Thomas, Tinker, Tojeiro, Tremonti, {Vargas Maga{\~{n}}a}, Verde,
  Viel, Wake, Watson, Weaver, Weinberg, Weiner, West, White, Wood-Vasey, Yeche,
  Zehavi, Zhao, \& Zheng}]{Dawson2013}
Dawson, K.~S., Schlegel, D.~J., Ahn, C.~P., {et~al.} 2013, Astronomical
  Journal, 145, 55, \dodoi{10.1088/0004-6256/145/1/10}

\bibitem[{DeBoor(1987)}]{deBoor1987}
DeBoor, C. 1987, {A practical Guide to Splines} (New York, NY: Springer)

\bibitem[{Demina {et~al.}(2016)Demina, Cheong, BenZvi, \&
  Hindrichs}]{Demina2016}
Demina, R., Cheong, S., BenZvi, S., \& Hindrichs, O. 2016, MNRAS, 480, 49,
  \dodoi{10.1093/mnras/sty1812}

\bibitem[{{DES Collaboration}(2005)}]{DES2005}
{DES Collaboration}. 2005, {The Dark Energy Survey: Status and First results},
  Tech. rep.

\bibitem[{Diggle(2013)}]{Diggle2013}
Diggle, P.~J. 2013, {Statistical analysis of spatial and spatio-temporal point
  patterns (3rd ed.)} (New York, NY: Chapman and Hall/CRC), 300,
  \dodoi{https://doi.org/10.1201/b15326}

\bibitem[{Dodelson \& Schneider(2013)}]{Dodelson2013}
Dodelson, S., \& Schneider, M.~D. 2013, Physical Review D - Particles, Fields,
  Gravitation and Cosmology, 88, \dodoi{10.1103/PhysRevD.88.063537}

\bibitem[{Durkalec {et~al.}(2018)Durkalec, {Le F{\`{e}}vre}, Pollo, Zamorani,
  Lemaux, Garilli, Bardelli, Hathi, Koekemoer, Pforr, \& Zucca}]{Durkalec2018}
Durkalec, A., {Le F{\`{e}}vre}, O.~L., Pollo, A., {et~al.} 2018, Astronomy and
  Astrophysics, 612, 1, \dodoi{10.1051/0004-6361/201730734}

\bibitem[{Eisenstein(2003)}]{Eisenstein2003}
Eisenstein, D.~J. 2003, The Astrophysical Journal, 586, 718,
  \dodoi{10.1086/367851}

\bibitem[{Eisenstein \& Hu(1997)}]{EisensteinHu1998}
Eisenstein, D.~J., \& Hu, W. 1997, The Astrophysical Journal, 496, 605,
  \dodoi{10.1086/305424}

\bibitem[{Eisenstein {et~al.}(2007)Eisenstein, Seo, Sirko, \&
  Spergel}]{Eisenstein2007}
Eisenstein, D.~J., Seo, H.-J., Sirko, E., \& Spergel, D.~N. 2007, The
  Astrophysical Journal, 664, 675, \dodoi{10.1086/518712}

\bibitem[{Eisenstein {et~al.}(2005)Eisenstein, Zehavi, Hogg, Scoccimarro,
  Blanton, Nichol, Scranton, Seo, Tegmark, Zheng, Anderson, Annis, Bahcall,
  Brinkmann, Burles, Castander, Connolly, Csabai, Doi, Fukugita, Frieman,
  Glazebrook, Gunn, Hendry, Hennessy, Ivezic, Kent, Knapp, Lin, Loh, Lupton,
  Margon, McKay, Meiksin, Munn, Pope, Richmond, Schlegel, Schneider, Shimasaku,
  Stoughton, Strauss, SubbaRao, Szalay, Szapudi, Tucker, Yanny, \&
  York}]{Eisenstein2005}
Eisenstein, D.~J., Zehavi, I., Hogg, D.~W., {et~al.} 2005, The Astrophysical
  Journal, 633, 560, \dodoi{10.1086/466512}

\bibitem[{Elvin-Poole {et~al.}(2017)Elvin-Poole, Crocce, Ross, Giannantonio,
  Rozo, Ryko, Avila, Banik, Blazek, Bridle, Cawthon, Friedrich, Kokron, Secco,
  Troxel, Abbott, Abdalla, Rosell, Carollo, Kind, Carretero, Castander, Cunha,
  Gruen, Gruendl, Gschwend, Gutierrez, Hartley, Hinton, Hoormann, Jain, James,
  Jarvis, Jeltema, Johnson, Percival, Petravick, Plazas, Romer, \&
  Sako}]{Elvin-Poole2017}
Elvin-Poole, J., Crocce, M., Ross, A.~J., {et~al.} 2017, Physical Review D, 98,
  042006, \dodoi{10.1103/PhysRevD.98.042006}

\bibitem[{Faltenbacher {et~al.}(2012)Faltenbacher, Li, \&
  Wang}]{Faltenbacher2012}
Faltenbacher, A., Li, C., \& Wang, J. 2012, Astrophysical Journal Letters, 751,
  \dodoi{10.1088/2041-8205/751/1/L2}

\bibitem[{Grimmett {et~al.}(2020)Grimmett, Mullaney, Bernhard, Harrison,
  Alexander, Stanley, Masoura, \& Walters}]{Grimmett2020}
Grimmett, L.~P., Mullaney, J.~R., Bernhard, E.~P., {et~al.} 2020, MNRAS, 000,
  1.
\newblock \doarXiv{2001.11573}

\bibitem[{Hamilton(1988)}]{Hamilton1988}
Hamilton, A. J.~S. 1988, The Astrophysical Journal, 331, L59,
  \dodoi{10.1086/185235}

\bibitem[{Hamilton(1993)}]{Hamilton1993}
---. 1993, Astrophysical Journal, 417, 19

\bibitem[{Hamilton \& Tegmark(2000)}]{Hamilton2000}
Hamilton, A. J.~S., \& Tegmark, M. 2000, Monthly Notices of the Royal
  Astronomical Society, 312, 285, \dodoi{10.1046/j.1365-8711.2000.03074.x}

\bibitem[{Hartlap {et~al.}(2007)Hartlap, Simon, \& Schneider}]{Hartlap2007}
Hartlap, J., Simon, P., \& Schneider, P. 2007, Astronomy and Astrophysics, 464,
  399, \dodoi{10.1051/0004-6361:20066170}

\bibitem[{Hatfield {et~al.}(2016)Hatfield, Lindsay, Jarvis, H{\"{a}}u{\ss}ler,
  Vaccari, \& Verma}]{Hatfield2016}
Hatfield, P.~W., Lindsay, S.~N., Jarvis, M.~J., {et~al.} 2016, Monthly Notices
  of the Royal Astronomical Society, 459, 2618, \dodoi{10.1093/mnras/stw769}

\bibitem[{Hawkins {et~al.}(2003)Hawkins, Maddox, Cole, Lahav, Madgwick,
  Norberg, Peacock, Baldry, Baugh, Bland-Hawthorn, Bridges, Cannon, Colless,
  Collins, Couch, Dalton, {De Propris}, Driver, Efstathiou, Ellis, Frenk,
  Glazebrook, Jackson, Jones, Lewis, Lumsden, Percival, Peterson, Sutherland,
  \& Taylor}]{Hawkins2003}
Hawkins, E., Maddox, S., Cole, S., {et~al.} 2003, Monthly Notices of the Royal
  Astronomical Society, 346, 78, \dodoi{10.1046/j.1365-2966.2003.07063.x}

\bibitem[{Hewett(1982)}]{Hewett1982}
Hewett, P.~C. 1982, Monthly Notices of the Royal Astronomical Society, 201,
  867, \dodoi{1982MNRAS.201..867H}

\bibitem[{Hinton {et~al.}(2019)Hinton, Howlett, \& Davis}]{Hinton2019}
Hinton, S.~R., Howlett, C., \& Davis, T.~M. 2019, MNRAS, 000, 1.
\newblock \doarXiv{1912.01175}

\bibitem[{Hogg(2008)}]{Hogg2008}
Hogg, D.~W. 2008, 1.
\newblock \doarXiv{0807.4820}

\bibitem[{Hu \& Sugiyama(1996)}]{HuSugiyama1996}
Hu, W., \& Sugiyama, N. 1996, The Astrophysical Journal, 471, 542,
  \dodoi{10.1086/177989}

\bibitem[{Illian {et~al.}(2008)Illian, Penttinen, Stoyan, \&
  Stoyan}]{Illian2008}
Illian, J., Penttinen, P.~A., Stoyan, H., \& Stoyan, D. 2008, {Statistical
  Analysis and Modelling of Spatial Point Patterns}, Statistics in Practice
  (Wiley).
\newblock \url{https://books.google.com/books?id={\_}U6BER2stYsC}

\bibitem[{Ivezic {et~al.}(2018)Ivezic, Kahn, {Anthony Tyson}, Abel, Acosta,
  Allsman, Alonso, AlSayyad, Anderson, Andrew, {Roger Angel}, Angeli, Ansari,
  Antilogus, Araujo, Armstrong, Arndt, Astier, Aubourg, Auza, Axelrod, Bard,
  Barr, Barrau, Bartlett, Bauer, Bauman, Baumont, Becker, Becla, Beldica,
  Bellavia, Bianco, Biswas, Blanc, Blazek, Blandford, Bloom, Bogart, Bond,
  Borgland, Borne, Bosch, Boutigny, Brackett, Bradshaw, {Nielsen Brandt},
  Brown, Bullock, Burchat, Burke, Cagnoli, Calabrese, Callahan, Callen,
  Chandrasekharan, Charles-Emerson, Chesley, Cheu, Chiang, Chiang, Chirino,
  Chow, Ciardi, Claver, Cohen-Tanugi, Cockrum, Coles, Connolly, Cook, Cooray,
  Covey, Cribbs, Cui, Cutri, Daly, Daniel, Daruich, Daubard, Daues, Dawson,
  Delgado, Dellapenna, de~Peyster, de~Val-Borro, Digel, Doherty, Dubois,
  Dubois-Felsmann, Durech, Economou, Eracleous, Ferguson, Figueroa,
  Fisher-Levine, Focke, Foss, Frank, Freemon, Gangler, Gawiser, Geary, Gee,
  Geha, {B Gessner}, Gibson, {Kirk Gilmore}, Glanzman, Glick, Goldina,
  Goldstein, Goodenow, Graham, Gressler, Gris, Guy, Guyonnet, Haller, Harris,
  Hascall, Haupt, Hernandez, Herrmann, Hileman, Hoblitt, Hodgson, Hogan, Huang,
  Huffer, Ingraham, Innes, Jacoby, Jain, Jammes, Jee, Jenness, Jernigan,
  Jevremovi, Johns, Johnson, {G Johnson}, {Lynne Jones}, Juramy-Gilles, Juri,
  Kalirai, Kallivayalil, Kalmbach, Kantor, Karst, Kasliwal, Kelly, Kessler,
  Kinnison, Kirkby, Knox, Kotov, Krabbendam, {Simon Krughoff}, Kub{\'{a}}nek,
  Kuczewski, Kulkarni, Ku, Kurita, Lage, Lambert, Lange, {Brian Langton}, {Le
  Guillou}, Levine, Liang, Lim, Lintott, Long, Lopez, Lotz, Lupton, Lust,
  MacArthur, Mahabal, Mandelbaum, Marsh, Marshall, Marshall, May, McKercher,
  McQueen, Meyers, Migliore, Miller, Mills, Miraval, Moeyens, Monet, Moniez,
  Monkewitz, Montgomery, Mueller, Muller, {Mu{\~{n}}oz Arancibia}, Neill,
  Newbry, Nief, Nomerotski, Nordby, Oliver, Olivier, Olsen, Ortiz, Osier, Owen,
  Pain, Palecek, Parejko, Parsons, Pease, {Matt Peterson}, Peterson, Petravick,
  {Libby Petrick}, Petry, Pierfederici, Pietrowicz, Pike, Pinto, Plante, Plate,
  Price, Prouza, Radeka, Rajagopal, Rasmussen, Regnault, Reil, Reiss, Reuter,
  Ridgway, Riot, Ritz, Robinson, Roby, Roodman, Rosing, Roucelle, Rumore,
  Russo, Saha, Sassolas, Schalk, Schellart, Schindler, Schmidt, Schneider,
  Schneider, Schoening, Schumacher, Schwamb, Sebag, Selvy, Sembroski, Seppala,
  Serio, Serrano, Shaw, Shipsey, Sick, Silvestri, Slater, {Allyn Smith}, {Chris
  Smith}, Sobhani, Soldahl, Storrie-Lombardi, Stover, Strauss, Street, Stubbs,
  Sullivan, Sweeney, Swinbank, Szalay, Takacs, Tether, Thaler, {Gregg Thayer},
  Thomas, Thukral, Tice, Trilling, Turri, {Van Berg}, {Vanden Berk}, Vetter,
  Virieux, Vucina, \& Wahl}]{Ivezic2018}
Ivezic, Z., Kahn, S.~M., {Anthony Tyson}, J., {et~al.} 2018, The Astrophysical
  Journal, 873, 44, \dodoi{10.3847/1538-4357/ab042c}

\bibitem[{Kaiser(2014)}]{Kaiser1987}
Kaiser, N. 2014, Monthly Notices of the Royal Astronomical Society, 227, 1,
  \dodoi{10.1093/mnras/227.1.1}

\bibitem[{Karagiannis {et~al.}(2014)Karagiannis, Shanks, \&
  Ross}]{Karagiannis2014}
Karagiannis, D., Shanks, T., \& Ross, N.~P. 2014, Monthly Notices of the Royal
  Astronomical Society, 441, 486, \dodoi{10.1093/mnras/stu590}

\bibitem[{Kazin {et~al.}(2012)Kazin, S{\'{a}}nchez, \& Blanton}]{Kazin2012}
Kazin, E.~A., S{\'{a}}nchez, A.~G., \& Blanton, M.~R. 2012, Monthly Notices of
  the Royal Astronomical Society, 419, 3223,
  \dodoi{10.1111/j.1365-2966.2011.19962.x}

\bibitem[{Kazin {et~al.}(2010)Kazin, Blanton, Scoccimarro, McBride, Berlind,
  Bahcall, Brinkmann, Czarapata, Frieman, Kent, Schneider, \&
  Szalay}]{Kazin2010}
Kazin, E.~A., Blanton, M.~R., Scoccimarro, R., {et~al.} 2010, Astrophysical
  Journal, 710, 1444, \dodoi{10.1088/0004-637X/710/2/1444}

\bibitem[{Kerscher(1999)}]{Kerscher1999}
Kerscher, M. 1999, Astronomy and Astrophysics, 343, 18.
\newblock \doarXiv{9811300}

\bibitem[{Kerscher {et~al.}(2000)Kerscher, Szapudi, \& Szalay}]{Kerscher2000}
Kerscher, M., Szapudi, I., \& Szalay, A. 2000, The Astrophysical Journal, 535,
  L13, \dodoi{10.1086/312702}

\bibitem[{Kipping(2010)}]{Kipping2010}
Kipping, D.~M. 2010, Monthly Notices of the Royal Astronomical Society, 408,
  1758, \dodoi{10.1111/j.1365-2966.2010.17242.x}

\bibitem[{Kitaura {et~al.}(2016)Kitaura, Rodr{\'{i}}guez-Torres, Chuang, Zhao,
  Prada, Gil-Marin, Guo, Yepes, Klypin, Sc{\'{o}}ccola, Tinker, McBride, Reid,
  S{\'{a}}nchez, Salazar-Albornoz, Grieb, Vargas-Magana, Cuesta, Neyrinck,
  Beutler, Comparat, Percival, \& Ross}]{Kitaura2016}
Kitaura, F.-S., Rodr{\'{i}}guez-Torres, S., Chuang, C.~H., {et~al.} 2016,
  Monthly Notices of the Royal Astronomical Society, 456, 4156,
  \dodoi{10.1093/mnras/stv2826}

\bibitem[{Landy \& Szalay(1993)}]{LandySzalay1993}
Landy, S.~D., \& Szalay, A.~S. 1993, The Astrophysical Journal, 412, 64

\bibitem[{Lanzuisi {et~al.}(2017)Lanzuisi, Delvecchio, Berta, Brusa, Comastri,
  Gilli, Gruppioni, Marchesi, Perna, Pozzi, Salvato, Symeonidis, Vignali, Vito,
  Volonteri, \& Zamorani}]{Lanzuisi2017}
Lanzuisi, G., Delvecchio, I., Berta, S., {et~al.} 2017, Astronomy and
  Astrophysics, 602, \dodoi{10.1051/0004-6361/201629955}

\bibitem[{Laureijs(2011)}]{Laureijs2011}
Laureijs, R.~J. 2011, {Euclid Definition Study Report}, Tech. rep.

\bibitem[{Li {et~al.}(2006)Li, Kauffmann, Jing, White, B{\"{o}}rner, \&
  Cheng}]{Li2006}
Li, C., Kauffmann, G., Jing, Y.~P., {et~al.} 2006, Monthly Notices of the Royal
  Astronomical Society, 368, 21, \dodoi{10.1111/j.1365-2966.2006.10066.x}

\bibitem[{Li {et~al.}(2016)Li, Park, Sabiu, Park, Weinberg, Schneider, Kim, \&
  Hong}]{Li2016}
Li, X.-D., Park, C., Sabiu, C.~G., {et~al.} 2016, The Astrophysical Journal,
  832, 1, \dodoi{10.3847/0004-637X/832/2/103}

\bibitem[{Mandelbaum(2018)}]{Mandelbaum2018a}
Mandelbaum, R. 2018, Annual Review of Astronomy and Astrophysics, 56, 393,
  \dodoi{10.1146/annurev-astro-081817-051928}

\bibitem[{Mukherjee \& Wandelt(2018)}]{MukherjeeWandelt2018}
Mukherjee, S., \& Wandelt, B.~D. 2018, Journal of Cosmology and Astroparticle
  Physics, \dodoi{10.1088/1475-7516/2018/01/042}

\bibitem[{Nuza {et~al.}(2013)Nuza, S{\'{a}}nchez, Prada, Klypin, Schlegel,
  Gottl{\"{o}}ber, Montero-Dorta, Manera, McBride, Ross, Angulo, Blanton,
  Bolton, Favole, Samushia, Montesano, Percival, Padmanabhan, Steinmetz,
  Tinker, Skibba, Schneider, Guo, Zehavi, Zheng, Bizyaev, Malanushenko,
  Malanushenko, Oravetz, Oravetz, \& Shelden}]{Nuza2013}
Nuza, S.~E., S{\'{a}}nchez, A.~G., Prada, F., {et~al.} 2013, Monthly Notices of
  the Royal Astronomical Society, 432, 743, \dodoi{10.1093/mnras/stt513}

\bibitem[{Padmanabhan {et~al.}(2006)Padmanabhan, White, \&
  Eisenstein}]{Padmanabhan2006}
Padmanabhan, N., White, M.~J., \& Eisenstein, D.~J. 2006, 000, 0,
  \dodoi{10.1111/j.1365-2966.2007.11554.x}

\bibitem[{Peebles \& Hauser(1974)}]{PeeblesHauser1974}
Peebles, P. J.~E., \& Hauser, M.~G. 1974, The Astrophysical Journal Supplement
  Series, 28, 19, \dodoi{10.1086/190308}

\bibitem[{Peebles \& Yu(1970)}]{PeeblesYu1970}
Peebles, P. J.~E., \& Yu, J.~T. 1970, The Astronomical Journal, 162, 815

\bibitem[{Percival {et~al.}(2014)Percival, Ross, S{\'{a}}nchez, Samushia,
  Burden, Crittenden, Cuesta, Magana, Manera, Beutler, Chuang, Eisenstein, Ho,
  McBride, Montesano, Padmanabhan, Reid, Saito, Schneider, Seo, Tojeiro, \&
  Weaver}]{Percival2014}
Percival, W.~J., Ross, A.~J., S{\'{a}}nchez, A.~G., {et~al.} 2014, Monthly
  Notices of the Royal Astronomical Society, 439, 2531,
  \dodoi{10.1093/mnras/stu112}

\bibitem[{Reid {et~al.}(2018)Reid, Seo, Leauthaud, Tinker, \& White}]{Reid2018}
Reid, B.~A., Seo, H.-J., Leauthaud, A., Tinker, J.~L., \& White, M.~J. 2018, {A
  2.5{\%} measurement of the growth rate from small-scale redshift space
  clustering of SDSS-III CMASS galaxies}, Tech. Rep. 0000.
\newblock \doarXiv{arXiv:1404.3742v2}

\bibitem[{Reid {et~al.}(2010)Reid, Percival, Eisenstein, Verde, Spergel,
  Skibba, Bahcall, Budavari, Frieman, Fukugita, Gott, Gunn, Ivezi{\'{c}},
  Knapp, Kron, Lupton, McKay, Meiksin, Nichol, Pope, Schlegel, Schneider,
  Stoughton, Strauss, Szalay, Tegmark, Vogeley, Weinberg, York, \&
  Zehavi}]{Reid2010}
Reid, B.~A., Percival, W.~J., Eisenstein, D.~J., {et~al.} 2010, Monthly Notices
  of the Royal Astronomical Society, 404, 60,
  \dodoi{10.1111/j.1365-2966.2010.16276.x}

\bibitem[{Riess {et~al.}(1998)Riess, Filippenko, Challis, Clocchiatti, Diercks,
  Garnavich, Gilliland, Hogan, Jha, Kirshner, Leibundgut, Phillips, Reiss,
  Schmidt, Schommer, Smith, Spyromilio, Stubbs, Suntzeff, \& Tonry}]{Riess1998}
Riess, A.~G., Filippenko, A.~V., Challis, P., {et~al.} 1998, The Astronomical
  Journal, 116, 1009, \dodoi{10.1086/300499}

\bibitem[{Ripley(1976)}]{Ripley1976}
Ripley, B.~D. 1976, Journal of Applied Probability, 13, 255,
  \dodoi{10.2307/3212829}

\bibitem[{Ripley(1981)}]{Ripley1981}
---. 1981, 252

\bibitem[{Rudemo(1982)}]{Rudemo1982}
Rudemo, R. 1982, Scandinavian Journal of Statistics, 9, 65

\bibitem[{S{\'{a}}nchez {et~al.}(2012)S{\'{a}}nchez, Sc{\'{o}}ccola, Ross,
  Percival, Manera, Montesano, Mazzalay, Cuesta, Eisenstein, Kazin, Mcbride,
  Mehta, Montero-Dorta, Padmanabhan, Prada, Rubi{\~{n}}o-Mart{\'{i}}n, Tojeiro,
  Xu, Maga{\~{n}}a, Aubourg, Bahcall, Bailey, Bizyaev, Bolton, Brewington,
  Brinkmann, Brownstein, Gott, Hamilton, Ho, Honscheid, Labatie, Malanushenko,
  Malanushenko, Maraston, Muna, Nichol, Oravetz, Pan, Ross, Roe, Reid,
  Schlegel, Shelden, Schneider, Simmons, Skibba, Snedden, Thomas, Tinker, Wake,
  Weaver, Weinberg, White, Zehavi, \& Zhao}]{Sanchez2012}
S{\'{a}}nchez, A.~G., Sc{\'{o}}ccola, C.~G., Ross, A.~J., {et~al.} 2012,
  Monthly Notices of the Royal Astronomical Society, 425, 415,
  \dodoi{10.1111/j.1365-2966.2012.21502.x}

\bibitem[{Satpathy {et~al.}(2016)Satpathy, Alam, Ho, White, Bahcall, Beutler,
  Brownstein, Chuang, Eisenstein, Grieb, Kitaura, Olmstead, Percival,
  Salazar-Albornoz, S{\'{a}}nchez, Seo, Thomas, Tinker, \&
  Tojeiro}]{Satpathy2016}
Satpathy, S., Alam, S., Ho, S., {et~al.} 2016, {The clustering of galaxies in
  the completed SDSS-III Baryon Oscillation Spectroscopic Survey: On the
  measurement of growth rate using galaxy correlation functions}, Tech. rep.
\newblock \doarXiv{1607.03148v2}

\bibitem[{Schneider \& Hartlap(2009)}]{Schneider2009}
Schneider, P., \& Hartlap, J. 2009, Astronomy and Astrophysics, 504, 705,
  \dodoi{10.1051/0004-6361/200912424}

\bibitem[{Sheth(2005)}]{Sheth2005}
Sheth, R.~K. 2005, Monthly Notices of the Royal Astronomical Society, 364, 796,
  \dodoi{10.1111/j.1365-2966.2005.09609.x}

\bibitem[{Sinha \& Garrison(2019)}]{Sinha2019}
Sinha, M., \& Garrison, L.~H. 2019, MNRAS, 000, 1.
\newblock \doarXiv{1911.03545}

\bibitem[{Skibba {et~al.}(2006)Skibba, Sheth, Connolly, \&
  Scranton}]{Skibba2006}
Skibba, R., Sheth, R.~K., Connolly, A.~J., \& Scranton, R. 2006, Monthly
  Notices of the Royal Astronomical Society, 369, 68,
  \dodoi{10.1111/j.1365-2966.2006.10196.x}

\bibitem[{Skibba {et~al.}(2014)Skibba, Smith, Coil, Moustakas, Aird, Blanton,
  Bray, Cool, Eisenstein, Mendez, Wong, \& Zhu}]{Skibba2014}
Skibba, R.~A., Smith, M. S.~M., Coil, A.~L., {et~al.} 2014, Astrophysical
  Journal, 784, \dodoi{10.1088/0004-637X/784/2/128}

\bibitem[{Sunyaev \& Zeldovich(1970)}]{SunyaevZeldovich1970}
Sunyaev, R., \& Zeldovich, Y. 1970, Astrophysics and Space Science, 7, 3

\bibitem[{Taylor \& Joachimi(2014)}]{TaylorJoachimi2014}
Taylor, A., \& Joachimi, B. 2014, Monthly Notices of the Royal Astronomical
  Society, 442, 2728, \dodoi{10.1093/mnras/stu996}

\bibitem[{Tessore(2018)}]{Tessore2018}
Tessore, N. 2018, arXiv, 5, \dodoi{10.3847/2515-5172/aad9a7}

\bibitem[{Trott {et~al.}(2019)Trott, Fu, Murray, Jordan, Line, Barry, Byrne,
  Hazelton, Hasegawa, Joseph, Kaneuji, Kubota, Li, Lynch, McKinley, Mitchell,
  Morales, Pindor, Pober, Rahimi, Takahashi, Tingay, Wayth, Webster, Wilensky,
  Wyithe, Yoshiura, Zheng, \& Walker}]{Trott2019}
Trott, C.~M., Fu, S.~C., Murray, S.~G., {et~al.} 2019, Monthly Notices of the
  Royal Astronomical Society, 486, 5766, \dodoi{10.1093/mnras/stz1207}

\bibitem[{Vargas-Maga{\~{n}}a {et~al.}(2013)Vargas-Maga{\~{n}}a, Bautista,
  Hamilton, Busca, Aubourg, Labatie, {Le Goff}, Escoffier, Manera, Mcbride,
  Schneider, \& Willmer}]{VargasMagana2013}
Vargas-Maga{\~{n}}a, M., Bautista, J.~E., Hamilton, J.-C., {et~al.} 2013,
  Astronomy {\&} Astrophysics, 554, A131,
  \dodoi{https://doi.org/10.1051/0004-6361/201220790}

\bibitem[{Wadekar {et~al.}(2020)Wadekar, Ivanov, \& Scoccimarro}]{Wadekar2020}
Wadekar, D., Ivanov, M.~M., \& Scoccimarro, R. 2020, 1.
\newblock \doarXiv{2009.00622}

\bibitem[{White \& Padmanabhan(2009)}]{WhitePadmanabhan2009}
White, M.~J., \& Padmanabhan, N. 2009, Mon. Not. R. Astron. Soc, 395, 2381,
  \dodoi{10.1111/j.1365-2966.2009.14732.x}

\bibitem[{Xu {et~al.}(2010)Xu, White, Padmanabhan, Eisenstein, Eckel, Mehta,
  Metchnik, Pinto, \& Seo}]{Xu2010}
Xu, X., White, M., Padmanabhan, N., {et~al.} 2010, Astrophysical Journal, 718,
  1224, \dodoi{10.1088/0004-637X/718/2/1224}

\bibitem[{Yuan {et~al.}(2020)Yuan, Jarvis, \& Wang}]{Yuan2020}
Yuan, Z., Jarvis, M.~J., \& Wang, J. 2020, The Astrophysical Journal Supplement
  Series, 248, 1, \dodoi{10.3847/1538-4365/ab855b}

\bibitem[{Zehavi {et~al.}(2011)Zehavi, Zheng, Weinberg, Blanton, Bahcall,
  Berlind, Brinkmann, Frieman, Gunn, Lupton, Nichol, Percival, Schneider,
  Skibba, Strauss, Tegmark, \& York}]{Zehavi2011}
Zehavi, I., Zheng, Z., Weinberg, D.~H., {et~al.} 2011, Astrophysical Journal,
  736, \dodoi{10.1088/0004-637X/736/1/59}

\end{thebibliography}

\end{document}